\documentclass[twocolumn]{aastex631}

\begin{document}

\title{The First Insights into an Ultraluminous X-ray Pulsar with XRISM: Phase-Resolved High-Resolution Spectroscopy of the Fe K-shell Band of M82 X-2}

\correspondingauthor{Shogo B. Kobayashi}

\author[0000-0001-7773-9266]{Shogo B. Kobayashi}
\email{shogo.kobayashi@rikkyo.ac.jp}
\affiliation{Department of Physics, Rikkyo University, 3-34-1 Nishi Ikebukuro, Toshima-ku, Tokyo 171-8501, Japan}

\author[0000-0003-4511-8427]{Peter Kosec}
\affiliation{Center for Astrophysics | Harvard \& Smithsonian, 60 Garden Street, Cambridge, MA 20138, USA}

\author[0009-0005-3295-7215]{Kazuki Ampuku}
\affiliation{Graduate School of Science, Nagoya University, Furo-cho, Chikusa-ku, Nagoya, Aichi, 464-8602, Japan}

\author[0000-0003-3244-0409]{Erin Boettcher}
\affiliation{Department of Astronomy, University of Maryland, College Park, MD 20742, USA}

\author[0000-0001-9894-295X]{Renata Cumbee}
\affiliation{NASA Goddard Space Flight Center, 8800 Greenbelt Rd., Greenbelt, MD 20771, USA}

\author[0000-0003-3462-8886]{Adam Foster}
\affiliation{Center for Astrophysics—Harvard-Smithsonian, Cambridge, MA 02138, USA }

\author[0000-0003-0058-9719]{Yutaka Fujita}
\affiliation{Department of Physics, Tokyo Metropolitan University, Tokyo 192-0397, Japan}

\author[0000-0001-8055-7113]{Kotaro Fukushima}
\affiliation{Institute of Space and Astronautical Science (ISAS), Japan Aerospace Exploration Agency (JAXA), Kanagawa 252-5210, Japan}

\author[0009-0001-2208-1310]{Skylar Grayson}
\affiliation{School of Earth and Space Exploration, Arizona State University, Tempe, Arizona, USA}

\author[0000-0003-3363-9786]{Gabriel Grell}
\affiliation{NASA Goddard Space Flight Center, 8800 Greenbelt Rd., Greenbelt, MD 20771, USA}

\author[0000-0002-2397-206X]{Edmund Hodges-Kluck}
\affiliation{NASA Goddard Space Flight Center, 8800 Greenbelt Rd., Greenbelt, MD 20771, USA}

\author[0000-0001-8667-2681]{Ann Hornschemeier}
\affiliation{NASA Goddard Space Flight Center, 8800 Greenbelt Rd., Greenbelt, MD 20771, USA}

\author[0009-0007-2283-3336]{Richard Kelley}
\affiliation{NASA Goddard Space Flight Center, 8800 Greenbelt Rd., Greenbelt, MD 20771, USA}

\author[0000-0001-9464-4103]{Caroline Kilbourne}
\affiliation{NASA Goddard Space Flight Center, 8800 Greenbelt Rd., Greenbelt, MD 20771, USA}

\author[0000-0002-1661-4029]{Mike Loewenstein}
\affiliation{Department of Astronomy, University of Maryland, College Park, MD 20742, USA}
\affiliation{NASA Goddard Space Flight Center, 8800 Greenbelt Rd., Greenbelt, MD 20771, USA}
\affiliation{Center for Research and Exploration in Space Science and Technology, NASA GSFC (CRESST II), Greenbelt, MD 20771, USA}

\author[0000-0002-9901-233X]{Ikuyuki Mitsuishi}
\affiliation{Department of Physics, Nagoya University, Aichi 464-8602, Japan}

\author[0000-0002-1875-6522]{Dustin Nguyen}
\affiliation{Ohio State University, 281 W Lane Ave, Columbus, OH 43210}

\author[0000-0002-3193-1196]{Evan Scannapieco}
\affiliation{School of Earth and Space Exploration, Arizona State University, P.O. Box 876004, Tempe, AZ 85287, USA}

\author[0000-0002-5504-4903]{Takeshi Tsuru}
\affiliation{Department of Physics, Kyoto University, Kyoto 606-8502, Japan }

\author[0000-0003-4885-5537]{Noriko Yamasaki}
\affiliation{Institute of Space and Astronautical Science (ISAS), Japan Aerospace Exploration Agency (JAXA), Kanagawa 252-5210, Japan}

\author[0000-0001-6366-3459]{Mihoko Yukita}
\affiliation{NASA Goddard Space Flight Center, 8800 Greenbelt Rd., Greenbelt, MD 20771, USA}
\affiliation{Johns Hopkins University, MD 21218, USA}



\begin{abstract}
During the performance verification phase, XRISM observed the M82 galaxy for a net exposure of 207.7 ks, with the ultraluminous X-ray pulsar (ULXP) X-2 included in the field of view. A pulsation search identified a candidate signal with a period close to the previously known value, 1.38727 s, at a significance of $3.15\sigma$ based on Monte Carlo simulations. Using this candidate period, phase-resolved spectral analysis with the high spectral resolution of Resolve was performed. The spectra suggest that, if the candidate pulsation is real, the Fe K$\alpha$ emission line in the pulse peak phase has a larger width ($36^{+60}_{-13}$ eV) than that in the remaining phase at a significance exceeding $3\sigma$. This suggests that at least a fraction of the Fe K$\alpha$ emission is associated with the ULXP system. The observed width corresponds to a velocity dispersion of $(1.7^{+2.8}_{-0.6})\times10^3$ km s$^{-1}$, which is too large to be explained by motions in the companion star atmosphere. The rise time of the pulsation constrains the line-emitting region to be smaller than $6.3\times10^4$ km, suggesting an origin in the accretion flow. This work demonstrates the capability of XRISM Resolve for pulsation-resolved high-resolution spectroscopy of ULX pulsars.
\end{abstract}



\section{Introduction} \label{sec:intro}
Ultra-luminous X-ray sources (ULXs; \citealt{makishima2000}) are non-nuclear point sources that emit X-rays with luminosity above the Eddington limit of the $10~M_{\rm\odot}$ black hole ($\sim10^{39}$~erg/sec). Although the nature of their central mass-accreting objects has been a mystery since their discovery \citep{fabbiano1989}, recent detections of X-ray pulsation from 8 ULXs (6 extragalactic and 2 Galactic) have revealed that some fractions of ULXs are neutron stars that accrete matter well above their Eddington rate. Since then, such ULX pulsars (ULXPs) have become one of the most ideal systems for studying the poorly understood mechanism of supercritical accretion.

Mass-accreting systems exhibit line features in X-ray spectra, reflecting the accretion environment/geometry surrounding the central objects. In particular, those from Fe, which have the highest fluorescence yield, are ubiquitously found and used to probe the velocity, ionization rate, and geometry of the accretion flows via their strength, central energy, and width. According to several numerical studies, the intense radiation pressure in the supercritical accretion flows inevitably forces the disk to launch a massive outflow (e.g., \citealt{ohsuga2006}), which can potentially form emission/absorption lines in the X-ray spectrum. Such studies have motivated researchers to search for any form of line, especially those from Fe, in the spectra of ULXs.

\citet{pinto2016} and \citet{pinto2021} have reported absorption and emission line features from highly ionized O, Ne, and N in the XMM-Newton reflection grating spectrometer data of several ULXs. The authors have found that the absorption lines are blue-shifted by $\sim20\%$ of the speed of light, indicating the presence of high-velocity outflows as theoretically expected. The systematic analyses of \citet{kosec2018a, kosec2021} show that soft X-ray ($0.3\--1.8$ keV) ULX spectra with sufficient S/N contain a range of emission and absorption lines. The emission lines, often from high-ionization N, O, and Ne transitions, are typically near their rest-frame energies; in contrast, the absorption lines appear to be far from rest, consistent with significant ($0.1\--0.2c$, where $c$ is the speed of light) blueshifts and an origin in high-velocity outflows. However, the findings were less definitive in the Fe features, particularly for the Fe emission lines.

\citet{walton2013} did not find evidence for Fe lines in Holmberg IX X-1 and established stringent upper limits on the equivalent width of possible undetected features of $15\--20$ eV. On the other hand, \citet{walton2016} detected an absorption line at 8.7 keV in NGC 1313 X-1, indicating an outflow with a blueshift of $0.2c$, consistent with the shift of the absorption lines detected in the soft X-ray band in the same object \citet{pinto2016}. \citet{kosec2018} found $3.7\sigma$ evidence for an outflow blueshifted by $0.22c$ in the ULXP NGC 300 ULX-1, imprinting high-ionization Fe absorption lines at $8\--9$ keV. \citet{brightman2022} detected an absorption line at 8.6 keV in the hyper-luminous ULX NGC 4045, which may indicate a high-velocity outflow or a cyclotron scattering feature. \citet{kobayashi2019} performed a stacking analysis of X-ray CCD data for representative extragalactic ULXs with the highest flux. The study established an upper limit of $<30$~eV for the strength of the Fe line features. The result suggests that, if present, the Fe line features must be weak and either too narrow or broad to be detected with the energy resolution and effective area of ordinary X-ray detectors.

Resolve \citep{porter2024, ishisaki2025, kelley2025} onboard the XRISM satellite \citep{tashiro2025} employs an X-ray microcalorimeter spectrometer that achieves a 5 eV energy resolution (FWHM) and a $>220$ cm$^2$ effective area at 6 keV, both greater than the Chandra HEG by a factor of 6. With these numbers combined, Resolve achieves nearly an order of magnitude better figure of merit for detecting weak lines around the Fe K-edge than the current best detector. After the launch on 2023/9/7, XRISM has observed 53 objects as part of its performance verification, including the M82 galaxy, which hosts several ULXs.

M82 is a starburst galaxy located at 3.5 Mpc \citep{dalcanton2009} and hosts an ULXP, X-2, in the near-core region, along with the brightest ULX, X-1, and several X-ray binaries within a $\sim20''$ radius. X-2 is the first ULX confirmed to be a pulser (a ULXP). Its pulsations were first detected with NuSTAR in 2014 \citep{bachetti2014}, during the observation of a type Ia supernova SN2014J \citep{goobar2014}. The spin period of the pulsar was initially $\sim1.32$ sec in 2001 and has shown a continuous spin-down trend over a 21-year timescale, down to $1.38$ sec \citep{liu2024}. Within an observation, however, X-2 has once been confirmed to be spinning up in the observation in 2014 \citep{bachetti2014}. Furthermore, the source has also exhibited other interesting phenomena, such as the disappearance of X-ray pulsation and period glitches \citep{bachetti2020}.  

Although the X-ray Mirror Assembly of the XRISM satellite has insufficient angular resolution ($\sim 1.3'$ in half-power diameter; \citealt{hayashi2024}) to resolve X-2 from other nearby X-ray sources, the time resolution of Resolve ($<100$ $\mu$s; \citealt{terada2025}) is adequate to detect the pulse period of $\sim 1.4$ sec, enabling us to distinguish the X-ray signals of X-2 from those of the surroundings. Thus, the performance verification observation has provided the first opportunity to search for Fe line features in ULXs that have been undetectable due to their narrow widths and weak intensities. In this paper, we report an analysis performed on the M82 observation data collected during the XRISM performance verification phase. All errors refer to a 68$\%$ confidence level unless stated otherwise.

\section{Observation and Data Reduction} \label{sec:obs}
The observation was conducted on 2024/05/12 (Observation ID: 300068010) with an effective net exposure of 207.7 ks. Although the data were taken at the end of an intense space storm in May, we included the data taken during this episode to maximize the statistics for the pulsation search. For the same reason, events from all pixels except for the calibration pixel are used in the analysis. Channel 27 is usually removed from analysis because its gain occasionally ``jumps'' for an unknown reason \citep{porter2024}, degrading the spectral energy resolution of that channel. However, there were no gain jumps during the observation, enabling us to utilize the channel in this study. The scripts used for the pre-pipeline and pipeline processes are version 005\_001.20Jun2024\_Build8.012 and 03.00.013.009, respectively. All calibrations, data reductions, and analyses are done using the software implemented in HEASoft version 6.34. 

We generated the response matrix file and the auxiliary response file with \texttt{rslmkrmf} and \texttt{xaarfgen}, respectively, following the procedures and parameters described in the XRISM Quick-Start Guide version 2.3, with the Calibration Database release 250315 in use. The size of the response matrix file is set to \texttt{large}. We analyze only the high-primary events, selecting those with {\tt GRADE=0:0} to maximize the energy resolution.
The arrival time of each event is corrected to the barycentric coordinate of the solar system using the JPL planetary ephemeris DE200 \citep{standish1990} and the M82 X-1 position $({\rm RA}, {\rm DEC}) = (148.98099^{\circ}, 69.678831^{\circ})$.

\section{Analysis and Result} \label{sec:result}
\subsection{Pulsation Search}
\label{sec:pulse_ana}
Since the pulsation of X-2 is known to be transient and its period is time-variable, we begin by searching for an X-ray pulsation and, if present, determining the best period candidate in this data set. According to previous studies, the pulse fraction of X-2 is generally $5\--10\%$ (e.g., \citealt{bachetti2014, liu2024}); less significant than that of other representative ULXPs (e.g., NGC 300 ULX-1: $\sim80\%$ at $8\--10$ keV \citealt{carpano2018}). Detecting such a weak pulsation requires a high-sensitivity searching method, and all of those in the previous searches for X-2 (e.g., \citealt{bachetti2014, bachetti2020, bachetti2021, bachetti2022, liu2024}) are based on the $Z^{2}_n$ statistic \citep{buccheri1983}, which is defined as
\begin{equation}
    Z^{2}_n = \frac{2}{N}\sum^{n}_{k=1}\left[\left(\sum^N_{j=1}\cos k\phi_j\right)^2 +\left(\sum^N_{j=1}\sin k\phi_j\right)^2 \right].
    \label{eq:z2}
\end{equation}
Here, $n$, $N$, and $\phi_j$ denote the number of harmonics, the total number of events, and the rotation phase of the $j$th event, respectively. 
The statistics are known to follow the $\chi^2$ distribution with $2n$ degrees of freedom. Hence, one can obtain the best pulsation period parameter and assess its statistical significance by searching for the parameter value that maximizes $Z^2_n$ in a given parameter space grid. The present study follows the same approach.

The rotation phase $\phi_j$ of the $j$th event, arrived at $t_j$, can be approximated using the rotation period $P_{\rm rot}$, its derivative $\dot{P}_{\rm rot}$, and the reference time $t_0$ as
\begin{equation}
    \phi_j = \frac{(t_{j}-t_0)}{P_{\rm rot}} -\frac{1}{2}\frac{(t_{j}-t_0)^2}{P_{\rm rot}^2\dot{P}_{\rm rot}} + \cdots.
    \label{eq:phi}
\end{equation}
If the object is a binary system, as in X-2, the arrival time should be corrected for Doppler modulation due to the binary orbital motion before calculating $\phi_j$. In the simplest binary model, namely a circular orbit binary with Newtonian approximation, this can be done by adding a factor $a\sin\theta_i\sin(2\pi t_{\rm ori}/P_{\rm orb}+\phi_0)/c$ to the arrival time $t_{\rm ori}$ of each X-ray photon. Here, $a$, $\theta_i$, $P_{\rm orb}$, and $\phi_0$ are the major axis radius, the orbital inclination angle, the binary orbital period, and the initial orbital phase at the reference time, respectively.

The pulse profile of X-2 has been roughly sinusoidal or single-peaked, and no second- or higher-order period derivatives have been detected in previous observations (e.g., \citealt{liu2024}). Hence, we assumed that the second- and higher-order terms in equation \ref{eq:phi} are zero and set $n=1$ for the pulsation search using equation \ref{eq:z2}. Although the binary orbit parameters are well constrained from the previous observations, \citet{bachetti2022} has found by following the orbital period for 7 years that the orbital period is not constant but exhibits a slight decay ($\dot{P}_{\rm orb}/P_{\rm orb}\sim -8\times10^{-6}$ year$^{-1}$) over time. Since the separation from the latest observation is nearly 3 years, we performed the pulsation search over the $P_{\rm orb}$, $a\sin\theta_{i}$, and $\phi_0$ space, in addition to $P_{\rm rot}$ and $\dot{P}_{\rm rot}$, within the uncertainties of the ephemeris and the decay rate derived by \citet{bachetti2022}. 

According to \citet{liu2024}, X-2 exhibited a continuous long-term spin-down trend from 2001 to 2022. The observation-to-observation spin-down rate had been approximately constant at $-7.4\times10^{11}$ Hz/s for a decade (from 2001/05/06 to 2011/04/09), and then gradually decreased in the subsequent observations. Assuming the recent trend has continued to the present, we may detect a pulsation by searching within a relatively narrow window around $\sim1.389$ s. However, when defining a search range, it is more appropriate to adopt the maximum measured spin-period derivative to date rather than extrapolating from the recent value. We therefore assumed a continuous $\pm1.298\times10^{-10}$ s/s spin-up/down rate, corresponding to the value observed between 2011 and 2022, from the nearest observation on 2021/10/18 ($P_{\rm orb}=1.387221$ s \citealt{bachetti2022}) to determine the upper and lower limits of the search range. Based on these assumptions, the search range for $P_{\rm rot}$ is set to $1.375774\-–1.398863$ s. As for $\dot{P}_{\rm rot}$ intrinsic to this observation, we scanned within $\dot{P}_{\rm rot}\le |5\times10^{-10}|~{\rm s~s}^{-1}$, which sufficiently covers the value range reported in the previous observations. 

The $P_{\rm orb}$ value at the present reference time (60443.50041 MJD) can be estimated as $P_{\rm orb}=2.532920$ days by extrapolating the orbital decay rate reported in \citet{bachetti2022}. Taking the uncertainties in the literature into account, we may conservatively set the scanning range of $P_{\rm orb}$ to $\pm10$ s around the estimated value. However, the combination of a $\sim250$ ks observational duration and a $\sim1.38$ s pulsation enables us to detect only a difference larger than $\sim300$ s in $P_{\rm orb}$, which is significantly longer than the scanning range. Hence, we concluded that the present observation is insensitive to a possible change in $P_{\rm orb}$ and fixed the value to the best estimate. The projected semi-major axis, $a \sin \theta_{i}$, was allowed to vary within a narrow interval around the reported value in \citet{bachetti2022} ($22.018\--22.318$ lt-s) to account for possible small systematic shifts. By modeling data points from 6 observations, \citet{bachetti2022} provided an empirical quadratic function that gives the time delay of the ascending node $\Delta T_{\rm asc}$ at a given observational time (Equation 4 in the literature). This can be used to estimate the initial orbital phase $\phi_0$ at the reference time of this observation, which becomes $\phi_0=0.464\pm0.004$. Because the model was derived from a limited number of observations spanning $56500\--59500$ MJD and the estimation above is an extrapolation, we allowed for additional systematic uncertainty when defining the search range. We searched the interval $\phi_0 = 0.464 \pm 0.024$, which is substantially larger than the statistical uncertainty derived from the quadratic fit and is intended to account for possible deviations from the quadratic orbital evolution.

To efficiently search for pulsations while keeping the computational cost manageable, we adopted a hierarchical search strategy. Instead of directly scanning the full parameter space at the resolution required for a fully coherent search, we first performed a coarse search using a sparse grid in the parameters $P_{\rm rot}$, $\dot{P}_{\rm rot}$, $a\sin\theta_i$, and $\phi_0$. Candidate signals identified in this stage were then re-examined with a finer grid in a second-stage coherent search.
The grid spacing for the fully coherent search is determined by the requirement that the phase mismatch due to parameter offsets should remain below $\sim1$ rad over the observation span $T_{\rm obs}\sim250$ ks. For $P_{\rm rot}$ and $\dot{P}_{\rm rot}$, the spacings required for the present observation are approximately $\sim7.7\times10^{-6}$ s and $\sim1\times10^{-11}~{\rm s~s^{-1}}$, respectively. As for the orbital parameters, the spacings in $a\sin\theta_i$ and $\phi_0$ are $\sim0.22$~lt-s and $\sim1.6\times10^{-3}$, respectively. In the first-stage coarse search, we set the grid size to be several times the full resolution to reduce computational cost while maintaining sufficient sensitivity to potential signals. The exception is $a\sin\theta_i$, whose scan range is already comparable to the required resolution, and we fixed it to $22.218$~lt-s. Specifically, the coarse grids for $P_{\rm rot}$, $\dot{P}_{\rm rot}$, and $\phi_0$ are set to $1.36\times10^{-5}~{\rm s}$, $3.3\times10^{-11}~{\rm s~s^{-1}}$, and $0.0069$, respectively.

The effective area drops sharply below 2 keV due to the gate-valve-closed configuration, and photon statistics are nearly zero even assuming the brightest state observed with Chandra \citep{brightman2020}. Therefore, we performed the coarse-to-fine search using data in $2\--12$ keV. The search using the coarse grid defined above yielded the maximum value $Z^{2}_{1}=34.6418$ at $P_{\rm rot}=1.38727~{\rm s}$, $\dot{P}_{\rm rot}=5\times10^{-10} ~{\rm s~s^{-1}}$, and $\phi_0 = 0.48$. Subsequently, we examined this candidate by re-scanning over a $\pm1$ coarse-grid window around the current-best parameters, using finer grids that satisfy the full resolution described in the previous paragraph. In this case, the number of grids and the scanning range for each parameter are as follows: 5 grids in $1.387256 ~{\rm s}\le P_{\rm rot}\le 1.387283~{\rm s}$, 6 grids in $4.667\times10^{-10}~{\rm s~s^{-1}}\le\dot{P}_{\rm rot}\le5.333\times10^{-10}~{\rm s~s^{-1}}$, 2 grids in $22.018$~lt-s $\le a\sin\theta_i \le$ $22.318$~lt-s, and 15 grids in $0.4731 \le\phi_0 \le 0.4868$. The second stage of the search exhibited the maximum statistic value of $Z^2_1=40.06$ at $P_{\rm rot}=1.3872699~{\rm s}$, $\dot{P}_{\rm rot}=4.8\times10^{-10} ~{\rm s~s^{-1}}$, $a\sin\theta_i=22.118~{\rm lt-s}$, and $\phi_0 = 0.4790$. We hereafter regard these as the best candidates for the pulsation search and proceed to evaluate their validity by taking a closer look at the $Z^2_1$ distribution around them.

\begin{figure}
    \centering
    \includegraphics[width=\columnwidth]{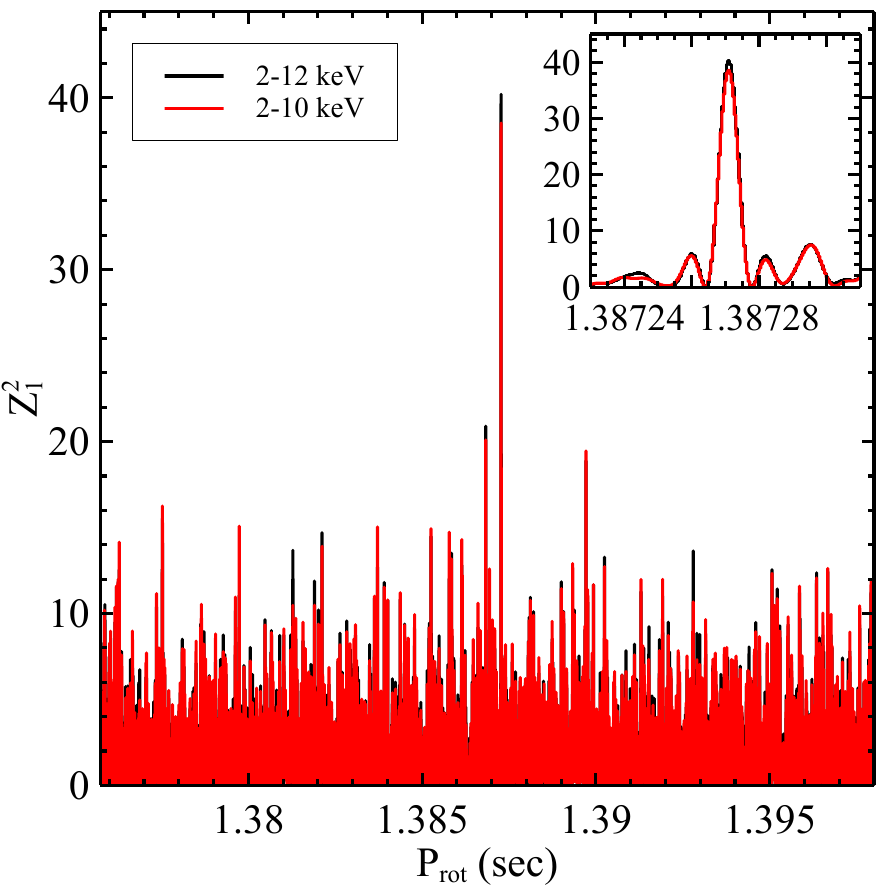}
    \caption{Periodogram at the best-estimate orbital parameter and period derivative. Black and red lines represent the result using $2\--12$ keV and $2\--10$ keV, respectively. The sub-panel is the same plot magnified around the peak.}
    \label{fig:periodgram}
\end{figure}
\begin{table}[]
    \centering
    \caption{The best parameters that give the highest $Z^{2}_{n}$}
    \begin{tabular}{lc}
    \hline
    \hline
       parameter  &  value \\
    \hline
    \hline
        $P_{\rm rot}$ & 1.387271(1) sec\\
        $\dot{P}_{\rm rot}$ & $4.8(3)\times10^{-10}$ sec sec$^{-1}$\\
        $a\sin\theta_i$ & 22.09(9) light sec\\
        $P_{\rm orb}$ & 2.53(3) day\\
        $\phi_0$ & 0.4789(6)\\
        reference time & 60443.50041 MJD \\
    \hline
    \end{tabular}
    \label{tab:psearch}
\end{table}
Figure \ref{fig:periodgram} shows the periodograms over the entire scan range of the rotation period. It is a projection of the $Z^2_1$ distribution at the best estimate of the orbital parameters and period derivative. The search using the full effective band of Resolve, namely $2\--12$ keV, yielded a peak at $P_{\rm rot}=1.387270$ s with {$Z^{2}_{1}=40.06$ (shown in black). The width of the peak is $\sim7\times10^{-6}~{\rm s}$ (see the sub-plot of Figure \ref{fig:periodgram}), which is comparable to the expected phase-mismatch resolution ($\sim7.7\times10^{-6}~{\rm s}$) of the coherent search. This consistency in width strengthens the argument that the potential modulation originated from the celestial signal. We also confirmed that the peak widths in other parameter spaces are similarly consistent with the expected resolution by creating 2D heat maps of $Z^2_1$ for all parameter combinations, as shown in Figure \ref{fig:search_result} of Appendix \ref{sec:p_search_significance}. Furthermore, each 2D map exhibits elliptical or ridge-like contour features, depending on the correlation between parameters. This is less likely to be observed if the possible modulation originates from statistical noise. Using these 2D maps, we finalized our best-estimate values and their uncertainties, as summarized in Table \ref{tab:psearch}.

Considering the gate-valve-closed configuration and the effective area of the X-ray mirror assembly, we also performed the same search in the $2\--10$ keV band to ensure that the peak is likely attributable to celestial signals by discarding energy bands that are potentially dominated by background events. The search still yielded a peak at a rotation period identical to that of the $2\--12$ keV case, with a maximum value of $Z^{2}_{1}=37.6$ (shown in red in Figure \ref{fig:periodgram}) under a decrease in the number of events from 66900 to 65398. 

Despite discarding the data in the $10\--12$ keV band, where background events are thought to dominate, the pulse significance has decreased slightly. To ensure that the background events are not the source of the possible modulation at $1.38727$ s, we also checked the $Z^{2}_{1}$ distribution of $15\--30$ keV, where the effective area of the X-ray mirror is known to be nearly zero for celestial photons. The $Z^2_1$ distribution of $15\--30$ keV did not show any peak-like structure at $(P_{\rm rot},~\dot{P}_{\rm rot})=(1.38727~{\rm sec}, 4.8\times10^{-10}~{\rm
sec~sec}^{-1})$, yielding a rather small value of $Z^2_1=4\--9$. In addition, we extracted non-X-ray background (NXB) data from the NXB database using \texttt{rslnxbgen} to compare it with the observational data. The extraction is performed by following the standard recipes prepared by the instrument team. The observed source count rate in $10\--12$ keV is $(9.6\pm0.2)\times10^{-3}$ counts s$^{-1}$, which is significantly higher than the NXB level, $(1.14\pm0.04)\times10^{-3}$ counts s$^{-1}$. This suggests that the data in $10\--12$ keV include some fractions of celestial events from X-2 and the M82 galaxy. On the other hand, the observed count rate in the $15\--30$ keV band is consistent with the NXB level, confirming that background events are dominant in this band.
Hence, the slight decrease in significance is likely due to discarding not only background signals but also celestial events in the respective energy band; the possible modulation is therefore unlikely to be attributed to the background events.

The obtained maximum value, $Z^2_1=40.06$, corresponds to a highly significant detection (chance probability of $1.9\times10^{-7}\%$). However, the true significance must account for the number of independent trials, which is generally difficult to estimate accurately. Therefore, we performed a Monte Carlo simulation to evaluate the probability of obtaining a higher $Z^2_1$ value than the observation by chance. Here, mock datasets equivalent to the observation but without coherent pulsation were generated by randomizing the arrival times of each event in the present data. We applied the identical (coarse-to-fine) search pipeline to 50000 mock datasets, and 80 of them yielded $Z^2_1$ higher than the observation. Hence, the false-alarm probability of the possible modulation is estimated as $0.16\%$, corresponding to a $3.15\sigma$ confidence level. Although this significance indicates that the pulsation detection should be regarded as tentative, the candidate periodicity provides a useful reference for phase-resolved spectroscopy. In particular, the first observation with a $\sim5$ eV energy resolution may reveal phase-dependent spectral variations that cannot be detected with previous instruments. We therefore proceed to further analysis using this candidate period, while noting that confirmation of the pulsation will require future observations.

\subsection{Pulse Profile}
\begin{figure}
    \centering
    \includegraphics[width=\columnwidth]{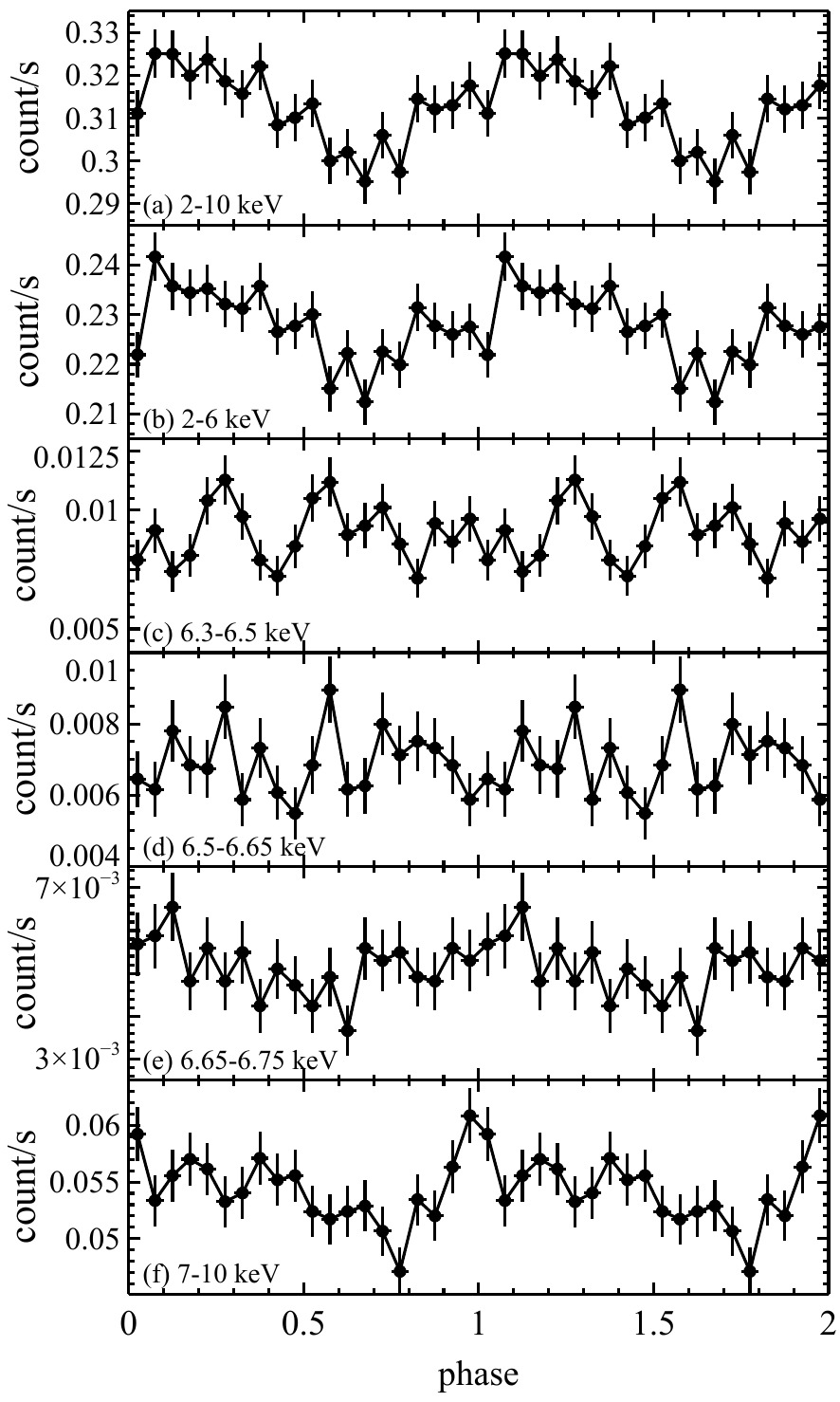}
    \caption{X-ray pulse profiles folded over the best parameters shown in Table \ref{tab:psearch}. An identical profile is presented twice for clarity. The errors in each data point correspond to the $68\%$ confidence level.}
    \label{fig:pprofile}
\end{figure}
Figure \ref{fig:pprofile} presents the folded pulse profiles from six representative energy bands: $2\--10$ keV (total energy band), $2\--6$ keV, $6.3\--6.5$ keV (the Fe K$\alpha$ line band), $6.5\--6.65$ keV, $6.65\--6.75$ keV (the Fe He$\alpha$ line band), and $7\--10$ keV. The pulse profile of the total energy band, namely $2\--10$ keV, is single-peaked and roughly sinusoidal. The pulsed fraction, defined as the count-rate ratio of the difference between the maximum and minimum over their total, increases with energy as $\sim7\%$ in the $2\--6$ keV band and $\sim13\%$ in the $7\--10$ keV band. These findings are consistent with previous studies (e.g., \citealt{bachetti2014, liu2024}). 

We examined the pulse profile in the $6.3\--6.5$ keV band by fitting it with a constant value, which yielded a reduced chi-squared of $\chi^2/{\rm dof}=33.8/19=1.78$, corresponding to a null-hypothesis probability of $\sim2\%$. Although this does not provide strong statistical significance, it suggests possible variability in the Fe K$\alpha$ band over the pulsation cycle. To evaluate the possible modulation using an independent method, we performed a $Z^2_n$ scan using events in the $6.3\--6.5$ keV band. No statistically significant peak was found in the $Z^2_n$ distribution near the best-estimate parameters. This is expected because the statistic is limited due to a narrow band, and the dual-peaked pulse profile, which is clearly not sinusoidal, yields a small $Z^2_n$ value under the $n=1$ assumption. In fact, increasing the harmonics to $n=5$ produces a weak peak in the $Z^2_n$ distribution at the same $P_{\rm rot}$ and $\dot{P}_{\rm rot}$ as those in Table \ref{tab:psearch}, with a maximum value of $Z^2_5=27$, although the statistical significance remains low. We also note that the count rate in $6.3\--6.5$ keV also includes contributions from continuum photons, raising the possibility that the observed hint of modulation is due to continuum contamination. Therefore, it may not be fully conservative to claim the detection of X-ray pulsations in the Fe K$\alpha$ line based solely on variability in the $6.3\--6.5$ keV count rate. Instead, a phase-resolved spectroscopy is required to confirm the true significance of the possible change in the Fe K$\alpha$ line, which will be described in the following sections.

Another result we found is that the pulse profile shows a hint of change with energy. While the profile of the total band ($2\--10$ keV) is roughly sinusoidal, that of the $7\--10$ keV band, which is likely dominated by X-ray continuum photons, is slightly asymmetric. It rises sharply within 20\% of the rotation period toward the peak at $\phi=0.0$ and then gradually decays over the remaining 70\% of the period. There may also be a hint of periodic subpeaks with a $\Delta\phi=0.2$ interval above the decaying trend. In addition, the pulse profile of the Fe K$\alpha$ band has two significant peaks at $\phi=0.3$ and $\phi=0.6$ (also a possible sub-peak at $\phi=0.7$). More interestingly, the first main peak of the Fe K$\alpha$ profile seems to lag behind the main peak of the $7\--10$ keV band, the photons in which contribute to generating the line emission, by $\Delta\phi=0.3$.

To evaluate the phase lag of the Fe K$\alpha$ band profile, we shifted the $7\--10$ keV pulse profile by a step of $\Delta\phi=0.05$ and calculated the cross-correlation factor between the $6.3\--6.5$ keV band. By repeating this 20 times, we scanned the entire rotation cycle and obtained the best lag candidate, which exhibits the largest correlation factor. Since the signal is periodic, the section that falls outside the phase range after the phase shift is recovered cyclically rather than discarded as in ordinary cross-correlation analysis. The best estimate is $\phi_{\rm lag}=0.3$, yielding a maximum correlation factor of $R=0.4$. The maximum correlation is rather moderate. This is because: 1) the pulse shapes share little similarity between the two bands, and 2) the peak-to-peak distance, which is $\Delta\phi=0.2$ in the $7\--10$ keV band and $\Delta\phi=0.3$ in the Fe K$\alpha$ band, is slightly different between the two bands. 

\subsection{Phase-resolved Spectral Analysis} \label{sec:spec_ana}
As the folded pulse profiles indicate variability in the Fe K$\alpha$ line and the continuum, we sliced the data by rotational phase and conducted a phase-resolved spectral analysis to confirm this. To quantify the variability, we fitted each spectrum using emission models implemented in XSPEC version 12.14. In the present spectral modeling, two collisional ionization equilibrium plasma models with photoelectric absorption (\texttt{tbabs*bapec+tbabs*bapec} in the XSPEC expression) are employed to account for thermal emissions from the surrounding diffuse gas that are irrelevant to X-2. The parameters for these thermal gas emissions are fixed to the values obtained in another study using the same dataset but focusing on the diffuse gas emission \citep{audard2026}. Since the appropriate physical model is poorly constrained, the overall continuum emissions from unresolved X-ray binaries, including X-2, are expressed empirically using a cutoff power-law model (\texttt{cutoffpl} in XSPEC), as in previous studies (e.g., \citealt{brightman2020}). Finally, the Fe K$\alpha$ emission is approximately reproduced by two Gaussians, each accounting for the K$\alpha_1$ and K$\alpha_2$ lines, respectively. The intensity ratio is fixed to $2:1$, and the width is tied between K$\alpha_1$ and K$\alpha_2$. Since the NXB intensity, $\sim6\times10^{-4}$ count/s/keV throughout $2\--10$ keV, is nearly one to one and a half orders of magnitude below the total spectrum, we consider its contribution negligible and fit the spectra with background included.

\begin{figure*}
    \centering
    \includegraphics[width=\linewidth]{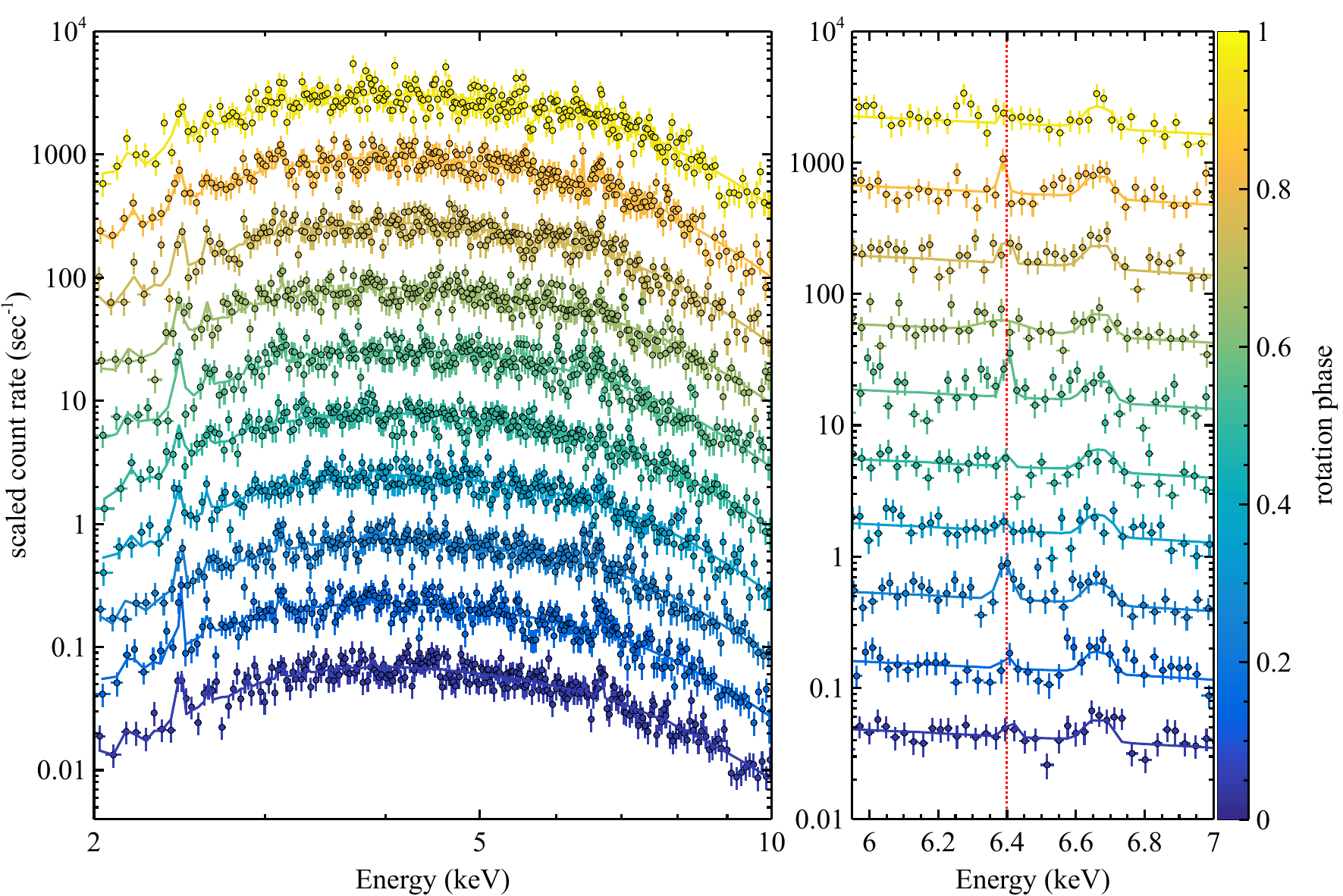}
    \caption{Left: the phase-resolved spectra extracted from each 0.1-rotational-phase bin. For clarity, the spectra are scaled vertically by a factor $3.33^{n}$, where $n=0, 1, 2,\cdots9$ is the number of bins separated from phase $0.0\--0.1$. The color of the spectra indicates the rotational phase from which it is extracted. The correspondence between the phase and color is shown in the color bar on the right. The solid lines represent the best-fit model in each spectrum. Right: same as the left panel but magnified around the Fe K$\alpha$ line. The red-dotted line indicates the rest-frame energy of the Fe K$\alpha$ line.}
    \label{fig:pres_spec}
\end{figure*}
Figure \ref{fig:pres_spec} presents the phase-resolved spectra extracted from each $\Delta\phi=0.1$ phase bin. Although the spectra exhibit nearly identical continua in the $2\--10$ keV band, the count rate around the Fe K$\alpha$ line shows a hint of variability, as expected from the pulse profile. In particular, the spectra from $\phi=0.2\--0.3$ and $\phi=0.5\--0.6$ seem to exhibit higher Fe K$\alpha$ intensities than the others, which is consistent with the phases where the count rate peaks are observed in Figure \ref{fig:pprofile} (c). The NXB count rate in $6.3\--6.5$ keV is estimated to be $(1.5\pm0.1)\times10^{-4}$ count/s from the stacked night Earth NXB spectrum extracted in Section \ref{sec:pulse_ana}. Thus, the background contribution in the Fe K$\alpha$ band is at most 2\%, which is too small to account for the $\sim\pm22\%$ count rate modulation.
The emission model has successfully reproduced each spectrum, and we have obtained the phase evolution of the model parameters, as shown in Figure \ref{fig:pres_spec}. For the fitting of $\phi=0.6\--0.7$, which has insufficient signal to constrain the Fe K$\alpha$ line Gaussian parameters, we fixed the line center at the rest-frame energy to obtain meaningful limits for the width and normalization. The details of parameter values and goodness-of-fit are summarized in Table \ref{tab:pres_fit}.
\begin{figure}
    \centering
    \includegraphics[width=\columnwidth]{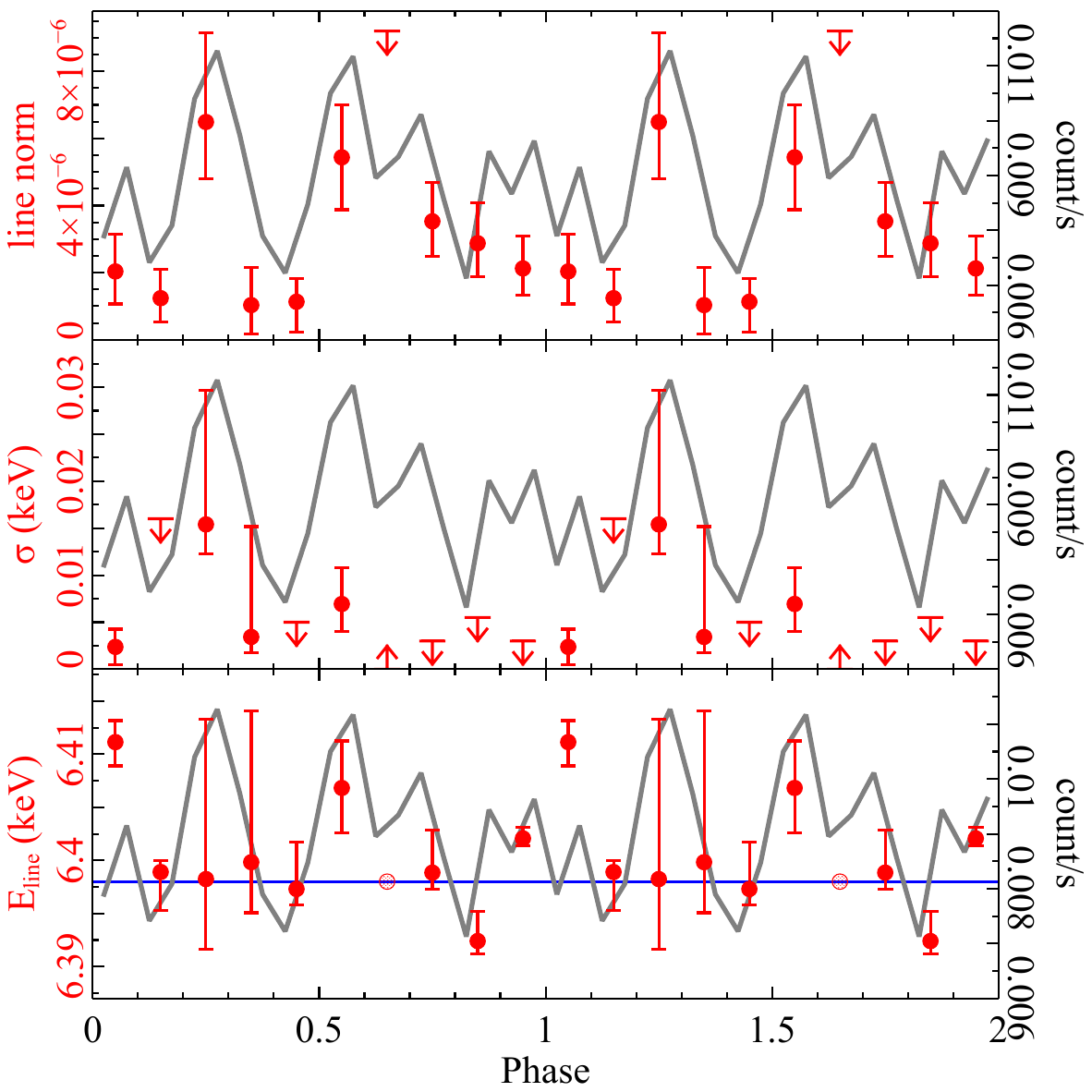}
    \caption{Phase evolution of the best-fit parameters derived from a spectral fitting of the phase-resolved spectra. The gray solid lines are the pulse profiles of the $6.3\--6.5$ keV band, which is the same as Figure \ref{fig:pprofile} (c) but without error bars. The horizontal line indicates the rest-frame energy assuming the redshift of the M82 galaxy ($z=0.000897$). The hatched data point at $\phi=0.6\--0.7$ denotes that the value is fixed at the rest-frame energy (see text for details).}
    \label{fig:par_profile}
\end{figure}

The normalization of the Fe K$\alpha$ line exhibits two peaks at $\phi=0.2\--0.3$ and $\phi=0.5\--0.6$, again at the same phases where the $6.3\--6.5$ keV count rate increases in the pulse profile. The $6.3\--6.5$ keV flux fractions of the continuum and Fe K$\alpha$ line are $0.84\--0.96$ and $0.04\--0.16$, respectively. If we assume that the $\sim5\%$ pulsed fraction of the $2\--10$ keV band is equally applicable to the continuum between $6.3\--6.5$ keV, then the estimated pulsed fraction contribution of the continuum in this energy band is $4.8\%$ at maximum, which is significantly smaller than the observed value of $\sim22\%$. Thus, the phase-resolved spectroscopy supports the presence of the varying Fe K$\alpha$ line. Furthermore, the line is likely broader ($10\--15$ eV) in these particular phase intervals, as the fits in the others mostly yielded upper limits of a few eV. We did not detect any shift in the line central energy at the $3\sigma$ level, except for the spectrum from $\phi=0.0\--0.1$. We must note, however, that the lower limit of the normalization reaches zero in this spectrum when we further calculate its uncertainty at the $3\sigma$ level, unlike the other phase bins that show high normalization values (e.g., $\phi=0.2\--0.3$). Therefore, we conclude that the intensity level of the Fe K$\alpha$ line at $\phi=0.0\--0.1$ is zero-consistent, and the possible line-centroid energy shift is due to statistical noise rather than a real feature.

\begin{table*}[]
    \caption{The best-fit parameters for the phase-resolved analysis.}
    \centering
    \begin{tabular}{lccccccc}
        \hline
        \hline
        phase &  $\Gamma^a$ & $kT_{\rm cut}$ & $F^c_{\rm cutoffpl}$ & $E^d_{\rm Fe_I}$ & $\sigma_{\rm Fe_I}$ & norm$^f_{\rm Fe_I}$ & C-stat/d.o.f \\
         & & (keV) & ($\times10^{-3}$) & (keV) & (eV) & ($\times 10^{-6}$) & \\
        \hline
        \hline
        $0.0\--0.1$ & $1.3\pm0.5$ & $>12$ & $4.6^{+0.4}_{-1.0}$ & $6.411\pm0.002$ & $2.4\pm1.9$ &  $2\pm1$ & 4118.8/5206\\
        $0.1\--0.2$ & $0.85\pm0.04$ &$7.1^{+2.0}_{-0.4}$ & $2.861^{+0.3}_{-0.005}$ & $6.399^{+0.001}_{-0.004}$ & $<16$ & $1.2^{+0.9}_{-0.7}$ & 4242.9/5223\\
        $0.2\--0.3$ & $0.4^{+0.4}_{-0.1}$ & $4.4^{+1.6}_{-0.5}$ & $2.0^{+0.9}_{-0.2}$ & $6.398^{+0.02}_{-0.007}$ & $15^{+14}_{-3}$ & $6^{+3}_{-2}$ & 4207.3/5193\\
        $0.3\--0.4$ & $0.20^{+0.5}_{-0.04}$ & $3.9^{+0.8}_{-0.2}$ & $1.811^{+0.195}_{-0.008}$ & $6.400^{+0.014}_{-0.005}$ & $3^{+1}_{-2}$ & $1.0^{+1.1}_{-0.9}$ & 4154.9/5127\\
        $0.4\--0.5$ & $1.20^{+0.04}_{-0.08}$ & $10.3^{+2.3}_{-0.8}$ & $4.0^{+0.5}_{-0.2}$ & $6.397^{+0.004}_{-0.001}$ & $<5$ & $1.1^{+0.7}_{-0.9}$ & 4145.3/5024\\
        $0.5\--0.6$ & $0.5\pm0.5$ & $4.7^{+0.2}_{-0.8}$ & $2.22^{+0.13}_{-0.01}$ & $6.407\pm0.004$ & $7^{+4}_{-3}$& $5\pm2$ & 3935.4/5075\\
        $0.6\--0.7$ & $0.7^{+0.6}_{-0.5}$ & $6^{+6}_{-2}$ & $2.4^{+0.2}_{-0.8}$ & $6.398$ (fixed) & - & $<9$ & 3956.4/4897\\
        $0.7\--0.8$ & $0.32\pm0.04$ & $4.1^{+0.7}_{-0.1}$ & $1.94^{+0.54}_{-0.07}$ & $6.399^{+0.004}_{-0.002}$ & $<3$ & $4\pm1$ & 3987.6/4930\\
        $0.8\--0.9$ & $0.44\pm0.04$ & $4.5\pm0.2$ & $2.194^{+0.536}_{-0.003}$ & $6.392^{+0.003}_{-0.001}$ & $<6$ & $3\pm1$ & 3971.7/5084\\
        $0.9\--1.0$ & $0.6\pm0.3$ & $5.7^{+1.4}_{-0.3}$ & $2.11^{+0.57}_{-0.09}$ & $6.402\pm0.001$ & $<3$ & $2.1^{+1.0}_{-0.8}$ & 4268.7/5108\\
        \hline
    \end{tabular}
    \tablecomments{a: The photon index of the \texttt{cutoffpl} model. b: The cutoff energy of the \texttt{cutoffpl} model. c: The flux of \texttt{cutoffpl} at 1 keV in units of photons/cm$^{2}$/sec/keV. d: The line-central energy of the Fe K$\alpha_1$ line. e: The $1\sigma$ width of Fe K$\alpha_1$ line. f: The XSPEC normalization of the Gaussian model representing the Fe K$\alpha_1$ line. For $\phi=0.6\--0.7$, the line width is unconstrained.}  
    \label{tab:pres_fit}
\end{table*}

\subsubsection{Fe K$\alpha$ On vs Off} \label{sec:fe_pulse_on_off_ana}
Since the Fe line normalization (and possibly the width) showed bimodal variability in Figure \ref{fig:par_profile}, we performed detailed spectroscopy by comparing spectra from the Fe-on with Fe-off phase. Here, we created a pair of spectra by co-adding the phase-resolved spectra shown in Figure \ref{fig:pres_spec} to increase the statistics and clarify the difference between them.
\begin{figure*}
    \centering
    \includegraphics[width=\linewidth]{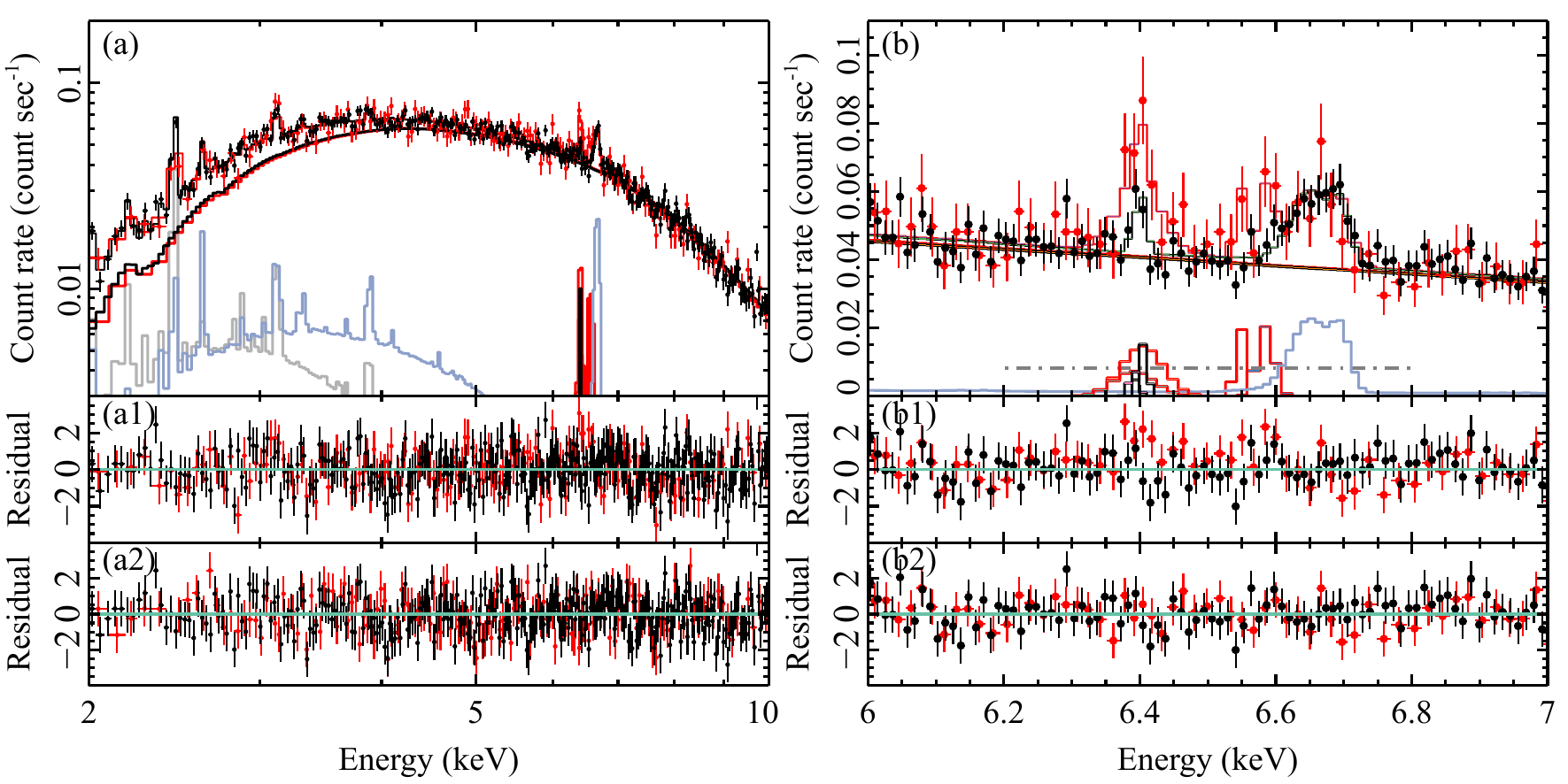}
    \caption{(a): Wide band Resolve spectra extracted from the pulse phase wherein the iron K$\alpha$ line is at its maximum (red) and the rest (black). Black and red solid lines represent the model related to the X-2 (including other point sources within the FOV) emission, namely the cutoff power-law and the Gaussian. The gray and light blue solid lines represent the persistent emission from the surrounding diffuse plasma. The bottom panels show the residuals from the best-fit model. (b) The same spectra as (a), but magnified around the iron K-shell emission. The dot-dashed line indicates the estimated contribution to the iron K$\alpha$ line from the surrounding diffuse emission, using Chandra observations (\citealt{iwasawa2023}; see text for details). Sub-panels: Residuals between the data and the model. Those shown in a1 and b1 are for the model assuming that the Fe K$\alpha$ intensity and width are the same in both spectra. Those in a2 and b2 correspond to the best-fitting model (see text for details).}
    \label{fig:fe_hilow_fit}
\end{figure*}
Figure \ref{fig:fe_hilow_fit} presents spectra taken from phase intervals $0.2<\phi<0.3$ and $0.5<\phi<0.6$, in which the Fe line intensity exhibited peaks (red), and another from the rest (black). In the spectral fitting, we used the $2\--10$ keV band of the data and jointly fitted the spectral pair with the same model as that in Section\ref{sec:spec_ana}. First, we tied all the Fe line parameters and allowed only the slope, intensity, and cutoff energy of \texttt{cutoffpl} to vary between the two. The model reproduced the wide-band spectra well, as shown in Figure \ref{fig:fe_hilow_fit} (a1). The continuum shapes turned out to be nearly identical between the two; however, the fluorescent Fe K$\alpha$ line clearly shows an intensity change over the X-ray pulsation (Figure \ref{fig:fe_hilow_fit}b). In fact, the model exhibited a positive residual around 6.4 keV, as shown in Figure \ref{fig:fe_hilow_fit} b1, indicating a change in the Fe K$\alpha$ line. Therefore, we untied all the Gaussian line parameters and refit the data.
\begin{figure}
    \centering
    \includegraphics[width=\columnwidth]{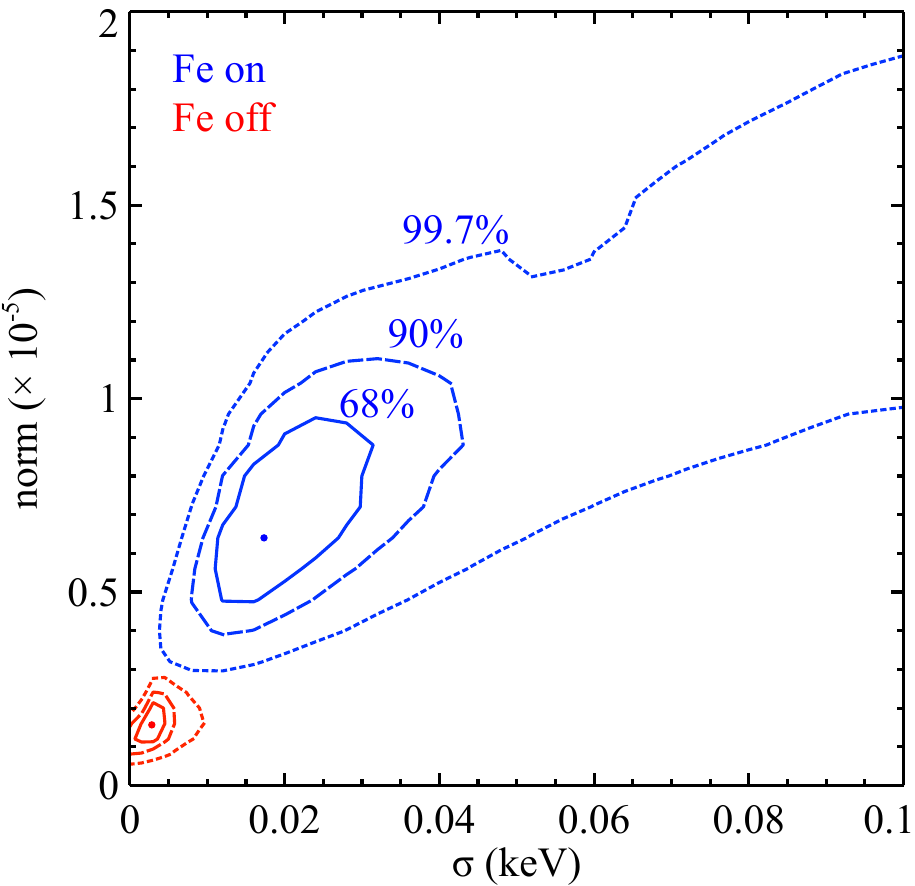}
    \caption{Significance contours of the width vs normalization of the Fe K$\alpha$ line. The on-pulse and the off-pulse results are shown in blue and red, respectively. The dots indicate the best-fit values. Toward the outside, each line corresponds to the significance of $68\%$, $90\%$, and $99.7\%$.}
    \label{fig:contour}
\end{figure}

\begin{table}[]
    \caption{The best-fit parameters for the Fe K$\alpha$ on/off group.}
    \centering
    \begin{tabular}{lcc}
        \hline
        \hline
        parameter &  on pulse & off pulse\\
        \hline
        \hline
        $\Gamma^a$ & $0.383\pm0.06$ & $0.573\pm0.06$\\
        $kT_{\rm cut}^b$ (keV) & $4.43\pm0.04$ & $5.16\pm0.04$\\
        $F_{\rm cutoff pl}^c~(\times10^{-3})$ & $2.07^{+0.02}_{-0.2}$ & $2.37^{+0.03}_{-0.08}$\\
        $E_{\rm Fe_{\rm I}}^d$ (keV) &$6.404\pm0.007$& $6.398\pm0.001$ \\
        $\sigma_{\rm Fe_{\rm I}}^e$ (eV) & $17^{+6}_{-4}$& $2.8^{+1.1}_{-0.9}$\\
        norm$_{\rm Fe_{\rm I}}^f$ ($\times10^{-6}$) & $6.4^{+3.2}_{-0.9}$& $1.6^{+0.4}_{-0.3}$\\
        C-stat/d.o.f & \multicolumn{2}{c}{19516/22339}\\
        \hline
    \end{tabular}
    \tablecomments{a: The photon index of the \texttt{cutoffpl} model. b: The cutoff energy of the \texttt{cutoffpl} model. c: The flux of \texttt{cutoffpl} at 1 keV in units of photons/cm$^{2}$/sec/keV. d: The line-central energy of the Fe K$\alpha_1$ line. e: The $1\sigma$ width of Fe K$\alpha_1$ line. f: The XSPEC normalization of the Gaussian model representing the Fe K$\alpha_1$ line.}  
    \label{tab:fe_group_fit}
\end{table}
Table \ref{tab:fe_group_fit} summarizes the best-fit parameters for this spectral group. Allowing the Gaussians to vary between the spectra has significantly improved the fit around the Fe K$\alpha$ line. The derived parameters suggest that the Fe K$\alpha$ line has a higher intensity and a larger width in the Fe band peak phase. As for the line center, no significant energy shift from the rest-frame value was detected. To confirm whether the change in width and intensity is statistically significant, we generated a confidence contour over the line width and normalization as shown in Figure \ref{fig:contour}. The contours do not overlap with more than a $3\sigma$ confidence level. Hence, we conclude that the intensity of the iron K$\alpha$ line varies over the pulsation phase, and that some fraction of its emission originates from X-2.

The contour also indicates that the K$\alpha$ line is narrower in the off-pulse phase. However, whether this narrower line is also attributed to X-2 is unclear. This is because a previous Chandra study has reported that Fe K$\alpha$ emission is present in the diffuse emission around the X-ray point sources in M82 \citep{iwasawa2023}. We estimated the potential contamination level of this diffuse K$\alpha$ emission in the present data by employing the observed intensity ratio between K$\alpha$ and He$\alpha$ lines in \citet{iwasawa2023}, as represented by the dot-dashed line in Figure \ref{fig:fe_hilow_fit} (b). The expected contamination intensity is roughly consistent with or marginally lower than the narrow line observed in the pulse-off phase. Hence, the off-pulse narrow line is likely dominated by emission from the gas surrounding the X-ray binaries, such as X-1 and X-2.

Although we could not rule out the possible contamination from the surrounding diffuse emission for the narrower Fe K$\alpha$ line in the pulse-off spectrum, we can safely conclude that at least the wider component in the pulse-on spectrum is attributed to the super-critically accreting pulsar in X-2. To estimate the net width and intensity of the line emission from X-2, we assumed that all of the contribution to the narrower Fe K$\alpha$ line originates from the surrounding diffuse emission, which is constant over the pulsation, and refitted the data by adding a pair of Gaussians (representing Fe K$\alpha_1$ and Fe K$\alpha_2$) whose parameters are coupled among the spectral pair. Under this assumption, the model exhibits a line width of $>0.1$ keV for the variable (wider) Fe K$\alpha$ line, which is significantly wider than the previous fit shown in Table \ref{tab:fe_group_fit}. This is because the count rate in several energy bins in the $6.5\--6.65$ keV band shows positive excesses, forcing the model to enlarge the width. Since the energies are consistent with the K$\alpha$ lines from Fe XXI (6.545 keV) and Fe XXII (6.585 keV), the origin of these bins may also be the pulsating component of X-2.

To examine whether the excessive data bins in $6.5\--6.65$ keV are signals associated with celestial emissions or statistical fluctuations, we first compared the pulse profile of this energy band with those from $6.3\--6.5$ keV (Fe K$\alpha$) and $6.6\--6.9$ keV (Fe He$\alpha$), as shown in panels (c) and (d) of Figure \ref{fig:pprofile}.
Although fitting the $6.5\--6.65$ keV pulse profile with a constant value yielded $\chi^2/{\rm dof} = 21.6/19$, which is statistically insufficient to reject the null hypothesis, the pulse profile exhibits hints of peaks at the same phase where the $6.3\--6.5$ keV band shows peaks. Furthermore, the pulse profile differs from that of the adjacent $6.5\--6.65$ keV band, which shows a minor degree of pulsation, as the signals are mostly dominated by Fe He$\alpha$ emission from the surrounding diffuse gas. Next, we added two Gaussians representing the Fe XXI and Fe XXII emission lines to the Fe-on spectral modeling. We examined how the fit improves compared to the previous result, which assumed only the Fe I K$\alpha$ line. The fit is performed assuming that Fe XXI and Fe XXII are in their rest frame. Namely, we fixed the line center energies to 6.545 and 6.585 keV, respectively. The fit improved from C-stat/dof = $19513$ to $19497$ ($\Delta C=16$) with 4 additional free parameters. Furthermore, the residuals around these lines have improved, as shown in Figures \ref{fig:fe_hilow_fit} (a2) and (b2). To evaluate the statistical significance of this improvement, we generated 10000 simulated spectra assuming the model without Fe XXI/XXII lines by using the \texttt{fakeit} command in XSPEC. The exposure of the dummy spectra is set to the same value as that of the observation. As a result, none of the spectra yielded $\Delta C > 16$, and the parameters were consistent within the errors after re-fitting them with additional Gaussians at 6.545 and 6.585 keV. Hence, the chance of obtaining a similar improvement to the observation due to random noise is smaller than $0.01\%$, and we conclude that the positive residuals in $6.55\--6.6$ keV are likely to be Fe line associated with the pulsating component.

Table \ref{tab:fe_group_refit} summarizes the best-fit parameters derived from the model with additional narrower Fe I, Fe XXI, and Fe XXII lines from those shown in Table \ref{tab:fe_group_fit}.
\begin{table}[]
    \caption{The re-fitting result for the Fe K$\alpha$ on/off group.}
    \centering
    \begin{tabular}{lcc}
        \hline
        \hline
        parameter &  on pulse & off pulse\\
        \hline
        \hline
        $\Gamma$ & $0.4\pm0.1$& $0.6^{+0.1}_{-0.2}$\\
        $kT_{\rm cut}$ (keV) & $4.4^{+0.6}_{-0.4}$& $5.25\pm0.5$\\
        $F_{\rm cutoff pl}~(\times10^{-3})$ & $2.1\pm0.4$ &$2.4^{+0.3}_{-0.2}$\\
        $E_{\rm Fe_{\rm I}~narrow}^a$ (keV) & $6.399\pm0.001$ & coupled\\
        $\sigma_{\rm Fe_{\rm I}~narrow}^b$ (eV) & $3.4^{+1.6}_{-0.9}$& coupled\\
        norm$_{\rm Fe_{\rm I}~narrow}^c$ & $1.6\pm0.4$& coupled\\
        $E_{\rm Fe_{\rm I}~wide}^d$ (keV)& $6.414^{+0.009}_{-0.016}$ & - \\
        $\sigma_{\rm Fe_{\rm I}~wide}^e$ (eV)& $36^{+60}_{-12}$& - \\
        norm$_{\rm Fe_{\rm I}~wide}^f$ ($\times10^{-6}$)& $7\pm3$& -\\
        $\sigma_{\rm Fe_{\rm XXI}}^g$ (eV)& $2\pm1$& -\\
        norm$_{\rm Fe_{\rm XXI}}^h$ ($\times10^{-6}$)& $2.1^{+0.9}_{-0.8}$ & -\\
        $\sigma_{\rm Fe_{\rm XXII}}^i$ (eV)& $12\pm7$ & -\\
        norm$^j_{\rm Fe_{\rm XXII}}$ ($\times10^{-6}$)& $4.0^{+1.5}_{-1.4}$ & -\\
        C-stat/d.o.f & \multicolumn{2}{c}{19497.32/22335}\\
        \hline
    \end{tabular}
    \tablecomments{a: The line-central energy of the Gaussian representing the narrower Fe I K$\alpha_1$ line. b: The $1\sigma$ width of the narrower Fe I K$\alpha_1$ line. c: The XSPEC normalization of the narrower Fe I K$\alpha_1$ line. d, e, and f: The same as a, b, and c, but for the broader Fe I K$\alpha_1$ line. g, h, i, and j: The same as b and c, respectively, but for the Fe XXI K$\alpha$ and the Fe XXII K$\alpha$ line.}  
    \label{tab:fe_group_refit}
\end{table}
Figure \ref{fig:fe_hilow_fit} shows the best-fit model (solid lines) and its residuals (panel b-2), respectively. Except for the width and normalization of the Fe lines, all derived parameters are consistent with those in Table \ref{tab:fe_group_fit} within the errors. Figure \ref{fig:contour_fexxi_fexxii} displays the confidence contour of the Fe I K$\alpha$ line width versus intensity. Adding lines from Fe XXI and Fe XXII has successfully prevented the fit from unnaturally broadening the broader (varying) Fe I K$\alpha$ line component and enabled us to constrain its width to $36^{+60}_{-13}$ eV at the $1\sigma$ confidence level. The width corresponds to a velocity dispersion of $(1.7^{+2.8}_{-0.6})\times10^3$ km/s. Furthermore, the contours do not reach zero at more than a $3\sigma$ level, indicating that a varying, broader Fe K$\alpha$ line is required on top of the narrow and constant one to reproduce the pulse-on spectrum.
\begin{figure}
    \centering
    \includegraphics[width=\columnwidth]{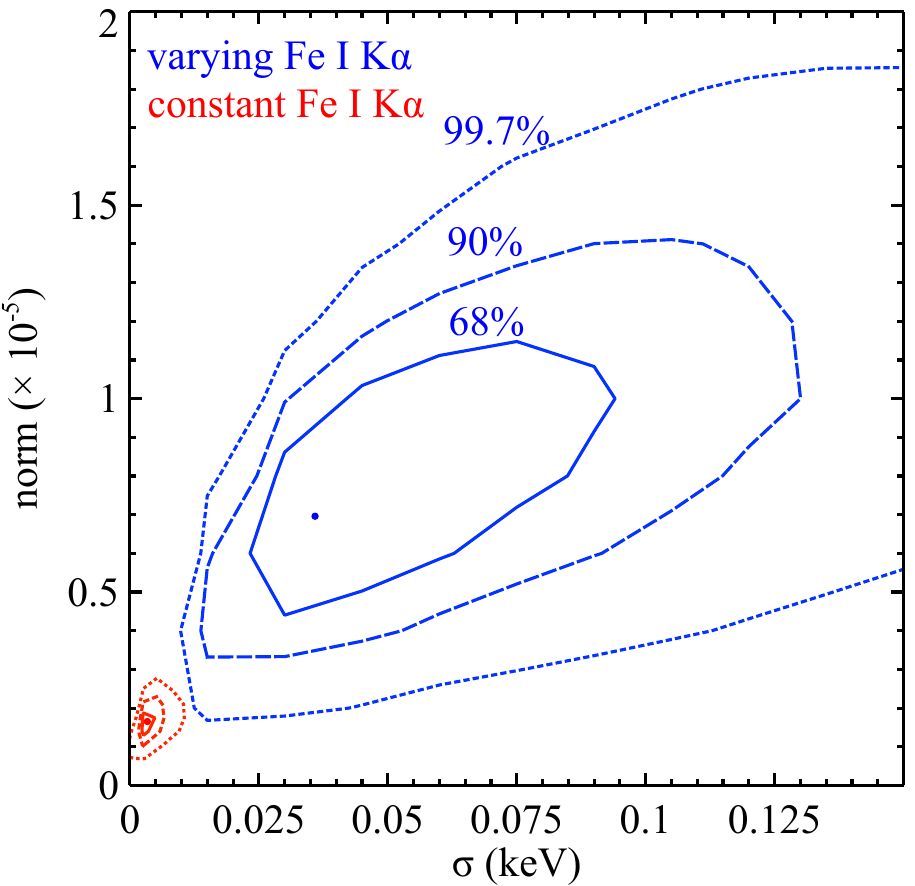}
    \caption{Significance contours of the width vs normalization of the Fe I K$\alpha$ lines for the model that includes the additional broadened (varying) Fe I K$\alpha$ line on top of the narrow (constant) one. The results for varying Fe I K$\alpha$ and a constant one are shown in blue and red, respectively. The dots indicate the best-fit value. Toward the outside, each line corresponds to the significance of $68\%$, $90\%$, and $99.7\%$.}
    \label{fig:contour_fexxi_fexxii}
\end{figure}
We must note, however, that this $>3\sigma$ confidence level for the line-width variability is conditional on the validity of the folding-period ephemeris. Therefore, the result is subject to the $3.15\sigma$ pulsation significance described in Section \ref{sec:pulse_ana}.

\subsubsection{Continuum Pulse On vs Off}
\begin{figure}
    \centering
    \includegraphics[width=\linewidth]{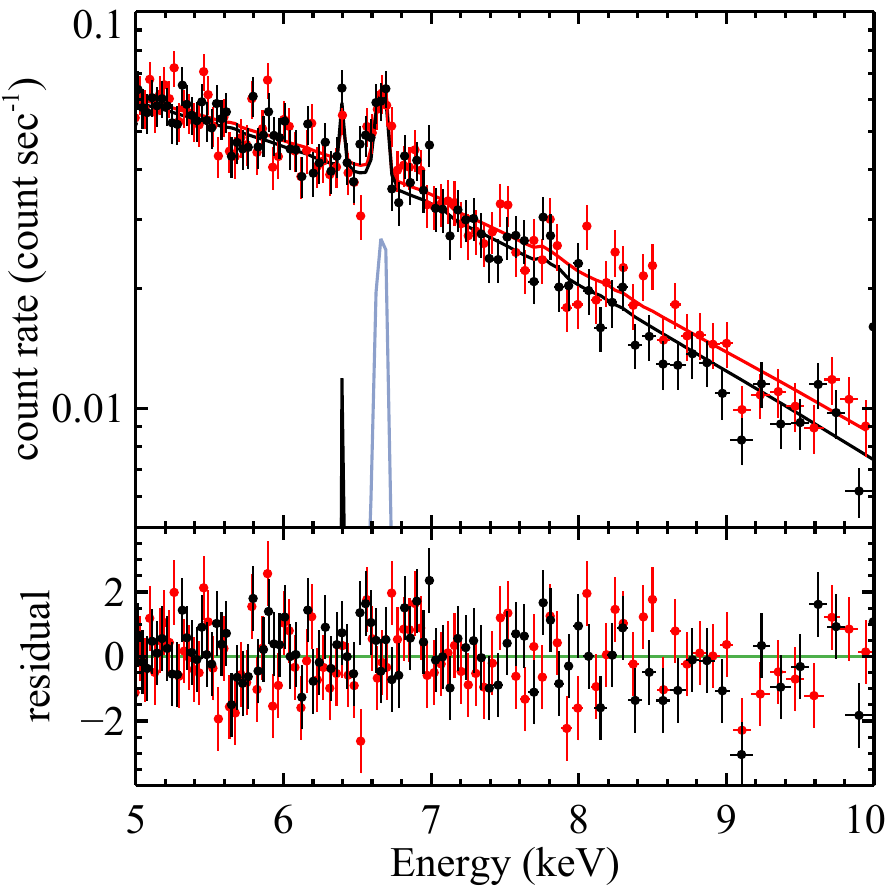}
    \caption{Resolve spectra extracted from the pulse phase wherein the $7\--10$ keV continuum intensity is at its maximum (red) and minimum (black). Black and red solid lines indicate the model for the X-2 emission: the cutoff power-law model and the Gaussian. The light blue solid lines represent the persistent emission from the surrounding diffuse plasma. The bottom panel shows the residuals from the best-fit model.}
    \label{fig:cont_high_low}
\end{figure}
In this section, we perform a phase-resolved spectral analysis based on the pulse shape of the $7\--10$ keV band to search for any variability in the continuum component. We again jointly fit the pair of spectra using the $2\--10$ keV band of the data and the same model as one initially employed in Section \ref{sec:fe_pulse_on_off_ana}. The top panel of Figure \ref{fig:cont_high_low} presents the continuum-on ($0.0 <\phi< 0.2$) and the continuum-off ($0.7<\phi<0.9$) spectra. In contrast to the spectra in Section \ref{sec:fe_pulse_on_off_ana}, the Fe line intensity is nearly consistent between the two spectra, whereas the continuum above 7 keV shows a hint of hardness change. To quantify this possible change in the spectral shape, we fit the spectra with the emission model, assuming the same configuration as that in Table \ref{tab:fe_group_fit}: constant optical plasma emissions from diffuse gas and variable cutoff power-law continuum and iron K$\alpha$ lines. The bottom panel of Figure \ref{fig:cont_high_low} shows the residuals from the best-fit model. The best-fit parameters are summarized in Table \ref{tab:continuum_fit}. Although some continuum shape parameters, such as the photon index and electron temperature, show hints of variability at 1$\sigma$ confidence level, we could not find any statistical difference in the spectral shape of M82 X-2.
\begin{table}[]
    \caption{The fitting result for the Continuum band group.}
    \centering
    \begin{tabular}{lcc}
        \hline
        \hline
        parameter &  on pulse & off pulse\\
        \hline
        \hline
        $\Gamma$ & $1.79^{+0.08}_{-0.13}$ & $1.5\pm0.2$\\
        $kT_{\rm cut}$ (keV) &  $>30$ & $12^{+9}_{-3}$\\
        $F_{\rm cutoff pl}$ ($\times10^{-3}$)& $7.2\pm0.6$ & $5.8^{+0.8}_{-0.6}$\\
        $E_{\rm Fe_{\rm I}}$ (keV) & $6.413^{+0.002}_{-0.003}$& $6.397^{+0.002}_{-0.001}$\\
        $\sigma_{\rm Fe_{\rm I}}$ (eV) & $8^{+7}_{-3}$ & $6^{+4}_{-2}$\\
        norm$_{\rm Fe_{\rm I}}$ ($\times10^{-6}$)& $1.7^{+1.0}_{-0.9}$ & $2.5^{+1.0}_{-0.9}$\\
        C-stat/d.o.f & \multicolumn{2}{c}{13020.08/16518}\\
        \hline
    \end{tabular}
    \label{tab:continuum_fit}
\end{table}

The residual in Figure \ref{fig:cont_high_low} shows a possible negative structure at $\sim 9$ keV, suggesting the presence of an absorption line. Suppose this is a blue-shifted Fe K-shell line; then the feature indicates an outflow with a velocity of $\sim0.24c$. However, if we perform the same simulation as in Section \ref{sec:fe_pulse_on_off_ana} (adding a Gaussian line at $\sim 9$ keV to the dummy spectra with no absorption line), the significance is at the $\sim 2\sigma$ level. Hence, we conclude that the feature is a possible hint of an outflow and recommend observing X-2 with additional exposure to test whether it is due to the stellar signal.

The difference in $2\--10$ keV flux between the on-pulse and the off-pulse is $\sim1.3\times10^{-12}$ erg sec$^{-1}$ cm$^{-2}$, which corresponds to a luminosity of $\sim2\times10^{39}$ erg sec$^{-1}$ assuming the distance to M82 is 3.5 Mpc. Although the total flux of X-2 is unknown due to severe contamination from other X-ray binaries, this result ensures that at least the pulsating component in X-2 itself was emitting X-rays at a $\sim10$ times the Eddington rate of the $1.4M_{\rm \odot}$ neutron star, and the total X-ray luminosity of the source is likely significantly higher. 

\section{Discussion and Conclusion} \label{sec:discussion}
\subsection{Long-term evolution of the pulse period \label{sec:period_evo}}

Although the detection significance is tentative, we overlaid our results with those from previous observations to present the long-term evolution of the pulse period in Figure \ref{fig:t_vs_p}.
\begin{figure}
    \centering
    \includegraphics[width=\columnwidth]{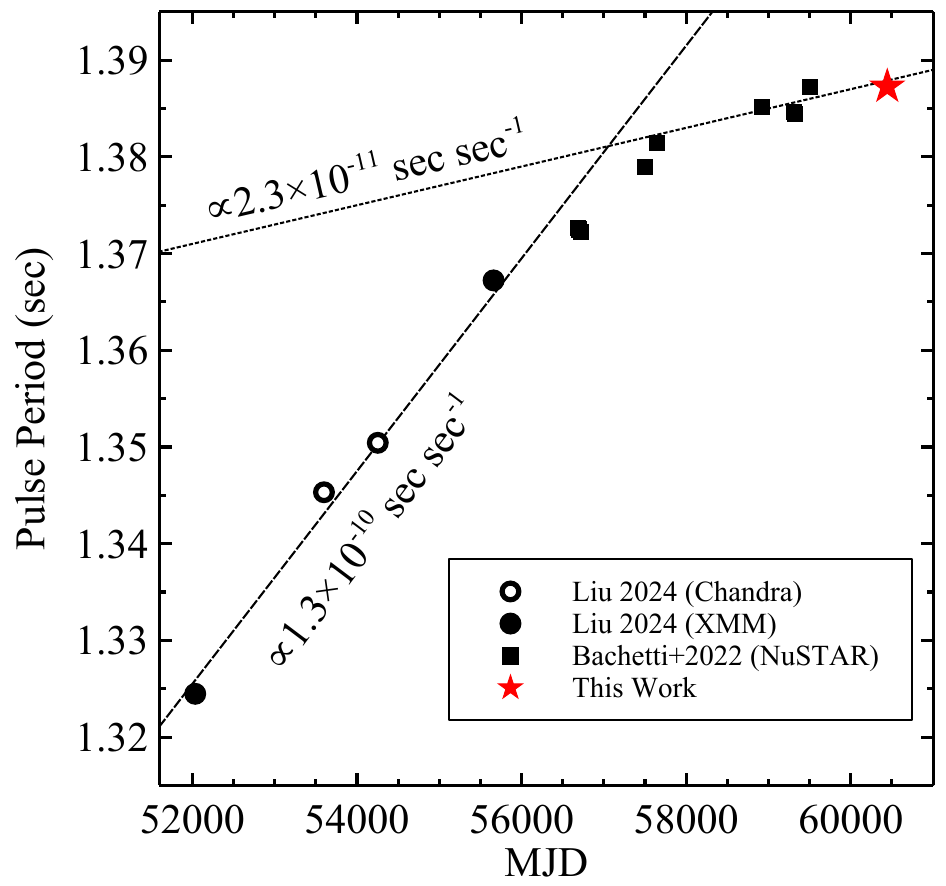}
    \caption{Long-term variability of the pulse period of M82 X-2. The dotted and dashed lines are eye guides indicating the respective spin-down rates.}
    \label{fig:t_vs_p}
\end{figure}
X-2 has been continuously spinning down since its first detection. As reported by \citet{liu2024}, the spin-down rate was rather constant at $\sim 10^{-10}$ sec/sec from the beginning up to 55660 MJD and has shown a tendency to saturate in the recent NuSTAR observations. The present result appears to be consistent with its extension. The spin period evolution from the previous observation by NuSTAR and XRISM is $1.387270-1.38722048=5\times10^{-5}$ sec, with a $\sim938$ day gap between them. This corresponds to a spin-down rate of $\sim6\times10^{-13}$ sec/sec; possibly, the source is reaching its spin equilibrium. This is also supported by the fact that the source alternates between spin-down and spin-up phases from one observation to the next, as described in \citet{bachetti2020}.

Mass-accreting neutron stars can acquire or lose angular momentum from the infalling gas via magnetic interaction at the magnetospheric radius $R_{\rm M}$, where the magnetic pressure of the neutron star's dipole magnetic field balances with the ram pressure of the accreting gas. Within this radius, the accreting gas is captured by the magnetic field line and transferred toward the neutron star's magnetic pole. Whether the neutron star spins up or down depends on the size relation between this $R_{\rm M}$ and the co-rotation radius $R_{\rm co}$. $R_{\rm co}$ is the radius where the Kepler velocity of accreting gas becomes equivalent to the neutron star rotation velocity, and is given as 
\begin{equation}
    R_{\rm co} = \left(\frac{P^2GM}{4\pi^2}\right)^{1/3}
    \label{eq:eq_rm}
\end{equation}
, where $G$, $M$, and $P$ are the gravitational constant, the neutron star mass, and the rotation period of the neutron star, respectively. The neutron star acquires angular momentum from accreting gas if $R_{\rm M} < R_{\rm co}$ is fulfilled. In contrast, the star can spin down if $R_{\rm M} > R_{\rm co}$ since the accreting gas carries away the angular momentum from the system as it is expelled by centrifugal force at $R_{\rm M}$. Accordingly, the spin-up and spin-down effects balance when $R_{\rm M}\sim R_{\rm co}$, thereby achieving spin equilibrium. If we assume that M82 X-2 is at its spin equilibrium, the magnetospheric radius can be directly obtained from the observed $P$ and equation \ref{eq:eq_rm} as $\sim2000$ km for a $1.4M_{\rm \odot}$ neutron star. Following the same method as \citet{bachetti2022}, we estimate the dipole magnetic field strength of M82 X-2 as $10^{13\--14}$ G (see Appendix E of \citealt{bachetti2022} for details), which is in a similar range to other estimations done in different ULXPs (e.g., \citealt{carpano2018}).

\subsection{Constraining the origin of the varying Fe lines}
Assuming that the pulsation, which is at a $3.15\sigma$ confidence level, is true, the high energy resolution of Resolve allows us to investigate possible phase-dependent variations of the Fe K$\alpha$ line in extragalactic ULXs with luminosities well above $10^{39}$ erg s$^{-1}$, i.e., about an order of magnitude higher than the Eddington limit for a $1.4 M_{\odot}$ neutron star. We emphasize that this discussion relies on the candidate pulsation detected in the $2\--12$ keV band, whose statistical significance remains tentative. Nevertheless, phase-resolved spectroscopy based on this candidate period suggests variability in the Fe K$\alpha$ emission. Even if the pulsation is not yet firmly established, the result demonstrates the new capability of XRISM Resolve for pulsation-resolved high-resolution spectroscopy of ULX pulsars. In the following, we discuss possible implications of the varying Fe K$\alpha$ line for the accretion geometry of M82 X-2.

Since the line intensity periodically varies with a neutron star rotation cycle (with a possible $\Delta\phi=0.3$ phase delay relative to the continuum), its origin is likely the pulse emission reprocessed somewhere in the accretion system surrounding the central neutron star. Three candidates are possible for the reprocessing region. 1) the stellar surface of the mass-donating star, 2) the stellar wind of the companion star, and 3) the accretion flow. In this section, we will estimate the most plausible origin among these candidates by utilizing the measured properties of the iron lines.

\subsubsection{Constraints from the line width \label{sec:limits_from_line_width}}
We have successfully constrained the width of the varying Fe K$\alpha$ line as $36^{+60}_{-13}$ eV, corresponding to a velocity dispersion of $(1.7^{+2.8}_{-0.6})\times10^3$ km/s. The velocity that creates this width can be roughly broken down into two components: binary motion and other intrinsic gas motions within the system. From the orbital parameters in Table \ref{tab:psearch}, the line-of-sight velocity due to the former can be calculated as $2\pi a\sin i/P_{\rm orb}\sim190$ km/s, corresponding to a width of $3.5$ eV. Hence, we cannot explain the observed value solely in terms of orbital motion, and additional velocity components from the three candidates are required to account for most of the remaining width. In this section, we test whether the obtained line width reconciles with each scenario listed above.

Let us first consider the possibility that the Fe K$\alpha$ line originates at the surface of the companion star. In this case, the line broadening can be attributed to the thermal and turbulent motions of the stellar atmosphere. Although the companion star of X-2 is unknown, previous studies (e.g., \citealt{bachetti2014}) have established a restriction that its mass must be $>5 M_{\rm \odot}$ (corresponding to a B or O-type star), considering its orbital parameters. According to the recent 2-D simulations, the atmospheric turbulence of the O-type star is estimated to be $30\--100$ km/s (e.g., \citealt{debnath2024}). In addition, the thermal broadening ($v/c=\sqrt{2/A}\times\sqrt{kT/m_{\rm p}c^2}$, where $v$, $A=56$, $T$, $k$, and $m_{\rm p}$ are the thermal velocity, atomic number, gas temperature, Boltzmann constant, and proton mass, respectively) at the stellar surface of the hottest ($T\sim4\times10^4$ K) O-type star is $\sim3$~km/s. Thus, both are significantly smaller than the observed value, and we conclude that the stellar surface is unlikely to be the origin of the Fe K$\alpha$ line.

In the case of the stellar wind origin, two possible velocity components can account for the line broadening: turbulent and bulk motion of the wind. The former velocity is estimated at $100\--200$ km/s (e.g., \citealt{runacres2002}), which is insufficient to account for the full width. The latter velocity, on the other hand, is typically $1000\--2000$ km/s for the O-type star wind (e.g., \citealt{prinja1990}), which is in the same order as the velocity dispersion inferred from the $\sim30$~eV line width, making it a possible candidate for the origin of the FeK$\alpha$ line. In that case, the line-emitting site should be a large area of the wind that covers a wide range of the line-of-sight velocity components to account for the line width.

Finally, we test the accretion-flow-origin scenario. In this scenario, the line width is likely attributed to gas motions in the accretion flow, such as the Keplerian motion. The present observed line energy width $\sigma_{E}$ can be approximately converted to the Kepler velocity $v_K$ using the inclination angle $i$ as
\begin{equation}
    v_K\sin i \sim \frac{\sqrt{2}}{2}\frac{c\sigma_{E}}{E_{\rm line}},
    \label{eq:keplarv}
\end{equation}
where $E_{\rm line}$ is the line center energy, namely 6400 eV. Substituting the obtained result, equation \ref{eq:keplarv} gives a projected-Kepler velocity of $(1.2^{+2.0}_{-0.4})\times10^3$ km/s. From the Kepler motion relation, $v_{\rm K}=\sqrt{GM/R}$, the distance of the line emitting region $R$ is estimated to be $(1.8^{+1.9}_{-1.5})\times10^5\sin^2 i$ km for a $1.4M_{\rm \odot}$ neutron star. Thus, assuming Keplerian motion, the line width provides an upper limit to the distance from the center toward the Fe line source.

\subsubsection{Constraints from the pulse profile \label{sec:lmits_from_pulse_profile}}
As shown in Figure \ref{fig:pprofile}, the Fe K$\alpha$ line band exhibits a dual-peaked pulse profile, suggesting that the Fe-line emitters are distributed locally along particular directions, and the illuminating beam has an opening angle comparable to the peak width. Each peak has a rising/falling time scale of $\sim 0.1$ and is separated from the other by $\sim 0.3$ of the rotation cycle. In terms of angles, these correspond to $\sim 18^{\circ}$ for the beam opening angle and $\sim 108^{\circ}$ for the distance between the two illuminated regions. Since the peak separation is not $180^{\circ}$, the bimodal profile is likely due to the actual position of the line-emitting areas rather than the illumination of a single spot by the opposite polar side of the neutron star. 

The rising and falling timescales of the pulse also impose a limit on the size of the Fe line emitter. The region illuminated by a pulse beam with an opening angle $\theta_0$ at a distance $R$ must have a size $D$ smaller than the light-crossing distance $cP\delta\phi = 4.1\times10^4$ km to catch up and vary its Fe fluorescence emission intensity in coincidence with the pulsation. Here, $\delta\phi=0.1$ and represents the phase duration of the rise/fall of the pulsation. The distance to the illuminated area is given as $R < 0.5D/\tan\theta_0$. If we substitute $\theta_0 = 18^{\circ}$ as the opening angle of the beam and $D=cP\delta\phi=4.1\times10^{4}$~km as the maximum size of the illuminated area, the maximum distance to the line-emitting source is given as $R=6.3\times10^4$ km.

\subsubsection{Constraints from the ionization parameter\label{sec:lmits_from_xi}}
As the observed line is the K$\alpha$ line from neutral Fe (and possibly from Fe XXII and XXI as well), the line-emitting region should be located at a large distance $r$ from the central ionizing source and/or have a density $n$ high enough to be less ionized under the emission exceeding the luminosity of $L_{\rm X}\sim2\times10^{39}$ erg sec$^{-1}$. In other words, the line-emitting region must have an ionization parameter, which is defined as $\xi=\frac{L_{\rm X}}{nR^2}$ \citep{tarter1969}, smaller than a certain value to emit the neutral K$\alpha$ line. Accordingly, the parameter provides us with restrictions over the distance vs density plane.

\begin{figure}
    \centering
    \includegraphics[width=\columnwidth]{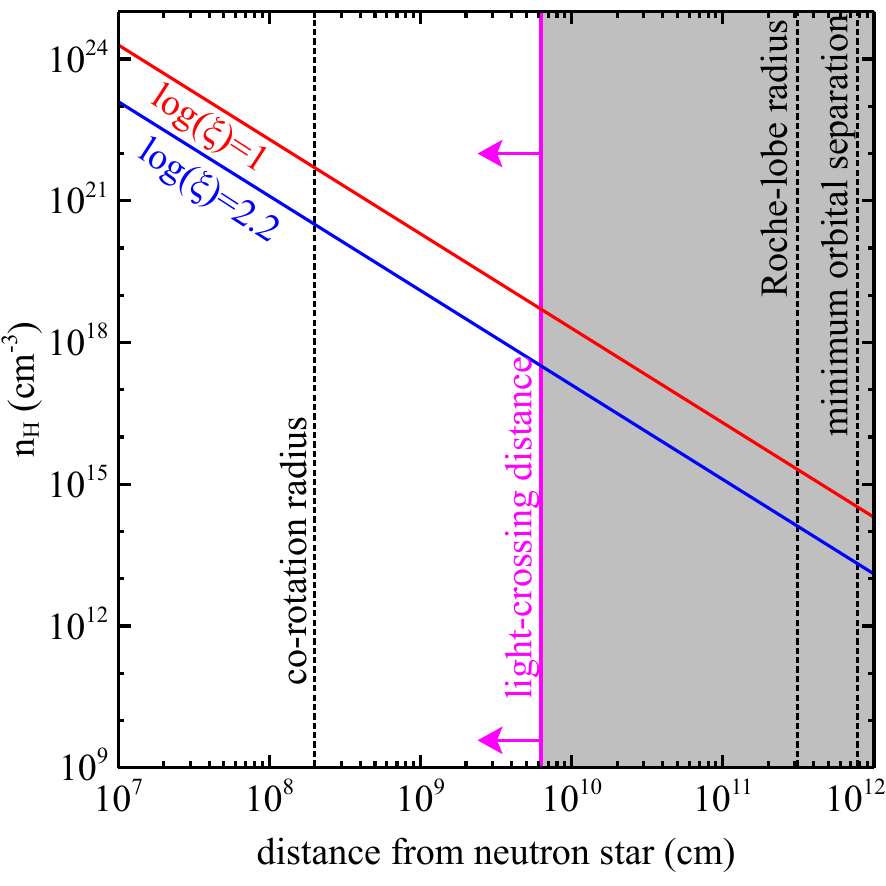}
    \caption{The relations of $\log(\xi)$ over the distance vs particle plane. The functions for $\log(\xi)=1$ (the condition to emit Fe I K$\alpha$ line) and $\log(\xi)=2.2$ (the condition to emit Fe XXI and Fe XXII K$\alpha$ lines) are shown in red and blue, respectively. The vertical lines indicate the distances toward the respective structures from the central neutron star. The gray-hatched area indicates the distance beyond which the Fe line emitter cannot be located, considering the light-crossing timescale (shown in magenta).}
    \label{fig:r_vs_n}
\end{figure}
 Figure \ref{fig:r_vs_n} presents the relation between the particle density $n$ and the distance from the central source $r$ over several ionization parameters $\xi$ (red and blue). We also overlaid the characteristic sizes obtained in Section \ref{sec:period_evo}, \ref{sec:limits_from_line_width}, and \ref{sec:lmits_from_pulse_profile}. According to \citet{kallman2004}, the gas must have $\log(\xi)$ smaller value than 1 for the neutral Fe K$\alpha$ line to be dominant ($\log(\xi)\sim 2.2$ for lines from Fe XXI and Fe XXII). Therefore, the gas emitting the Fe K$\alpha$ line must be more distant and/or denser than the curve shown in red in Figure \ref{fig:r_vs_n}. In addition to this limitation, the light-crossing distance derived in Section \ref{sec:lmits_from_pulse_profile} provides a stringent restriction on the distance (the magenta line). Combining these two, the line emitting region is required to be within $6.9\times10^4$ km from the central source, which is significantly smaller than the neutron star's Roche-lobe radius, and to have a density as large as $10^{18}\--10^{23}$ cm$^{-3}$. As the O-type star stellar wind is estimated to have a maximum density of $\sim10^{13}$ cm$^{-3}$ at the stellar surface (e.g., \citealt{debnath2024}), this condition is too close and dense for stellar wind. Hence, we conclude that the origin of the pulsating Fe K$\alpha$ line is unlikely to be the stellar wind, and a dense accretion flow, such as a thin accretion disk (typical density higher than $10^{20}$ cm$^{-3}$), in the vicinity of the neutron star is the best candidate.

\subsection{A possible accretion geometry of M82 X-2}
Finally, let us propose a possible accretion geometry that can fulfill the conditions derived in the previous discussion using the Fe K$\alpha$ line. The conditions obtained so far are as follows. (1) The pulse profile of the continuum band has a single-peaked shape, whereas that of the Fe K$\alpha$ band is dual-peaked. (2) The peak-to-peak distance in the Fe K$\alpha$ band pulse profile is $\sim108^{\circ}$, suggesting that the illuminated regions are localized to those particular angles. (3) Considering the light-crossing distance and the ionization state, the Fe-line emitting source should have a size smaller than $4.1\times10^{4}$ km and be located at a distance closer than $7\times10^{4}$ km, with a particle density of $10^{18\--23}$ cm$^{-3}$ or higher. (4) The gas emitting the Fe-K$\alpha$ line must have a sufficient range in its velocity distribution to account for the observed line width of $36^{+60}_{-13}$ eV, which corresponds to $(1.2^{+2.0}_{-0.4})\times10^3 \sin i$ km/s, assuming the Kepler velocity. 

As previously described, conditions (3) and (4) disfavor the stellar surface or stellar wind being the origin of the Fe K$\alpha$ line, and the current best candidate is the accretion flow around the neutron star. In an accreting neutron star system, the infalling gas is expected to form an accretion disk around the central object. If the neutron star is sufficiently magnetized, the magnetic field truncates the accretion disk at $R_{\rm M}$, and the gas within the radius is channeled onto the magnetic pole via magnetic field lines. The concentrated gas forms a cylindrical structure above the magnetic pole, known as the accretion column (e.g., \citealt{basko1975, basko1976}), in which the shock-heated gas generates high-energy X-ray photons. Under high-accretion-rate conditions, the accretion column becomes optically thick, and photons escape from its sidewalls, creating an emission pattern perpendicular to the magnetic field, called a fan beam \citep{basko1976}. In addition, some fractions of the fan beam photons may irradiate the neutron star's surface near the magnetic pole and be reflected as an alternate emission pattern that is relatively collimated and parallel to the magnetic axis, known as a polar beam or pencil beam (e.g., \citealt{trumper2013, poutanen2013}). These two major beam components can form a pulsating emission pattern and periodically illuminate the surrounding accretion structure as the entire accretion column rotates with a given offset angle from the neutron star rotation axis. As the Fe K$\alpha$ band has a sharper profile than the continuum (condition 1), the polar beam, which has a narrower opening angle, is likely illuminating the surrounding structure, while the fan beam emission reaches the observer directly. 

Considering the solid angle from the central pulsating emission and required density as high as $10^{18\--23}$ cm$^{-3}$ (condition 3), one of the best candidates for the Fe line emission site is the inner wall of the accretion disk truncated at $R_{\rm M}$. Hence, we subsequently assume this inner-disk-wall scenario as a working hypothesis and test whether it can realize the observational conditions listed above. \citet{bykov2022} and \citet{xiao2024} reported such an example in the observation of the Galactic ULXP SWIFT J0243.6+6124. Theoretical studies suggest that, in super-critical accretion flows, radiation pressure overcomes the gravitational pull of the central object within a certain radius called the spherization radius $R_{\rm sp}$ (e.g., \citealt{shakura1973, ohsuga2006, poutanen2007}) and forces the disk to become geometrically thick. Hence, the compact object can be surrounded by a wall of ``well'' formed by the inflated accretion disks (e.g., \citealt{kawashima2012}) in a super-critical condition. \citet{bykov2022} and \citet{xiao2024} concluded that the pulsating Fe line in SWIFT J0243.6+6124 could be explained as a result of central beam emission being reflected at such a thick wall of the inner disk region. 

The spherization radius is approximately given as $R_{\rm sp}\sim 3R_{\rm s}\dot{m}$ (e.g., \citealt{shakura1973, poutanen2007}), where $R_{\rm s}$ and $\dot{m}$ are the Schwarzschild radius and the mass-accretion-rate ratio over the Eddington rate, respectively. If we assume that the entire X-ray flux in $2\--10$ keV ($\sim2\times10^{-11}$ erg sec$^{-1}$ cm$^{-2}$) is from X-2, namely assuming the maximum luminosity ($L_{\rm X} \sim 3\times10^{40}$ erg sec$^{-1}$), and the mass accretion rate $\dot{M}$ is given via luminosity as $\dot{M} = L_{\rm X}R/(GM)$, the estimated spherization radius in X-2 is at most $R_{\rm sp}\sim10^{3}$ km. Thus, $R_{\rm sp}$ in X-2 is significantly smaller than the $R_{\rm M}\sim 2000$ km estimated in Section \ref{sec:period_evo} using the co-rotation radius, suggesting that the disk is truncated by the magnetic field before being geometrically thick due to the radiation pressure. This is in favor of conditions (1) and (3) because a thin disk tends to have a higher density and is therefore less ionized than an inflated one under the same accretion rate. In addition, it may explain the relatively sharp pulse profile of the Fe K$\alpha$ line band as the small scale height limits the time duration of the central pulse beam illuminating the accretion disk.

In the inner disk-origin scenario, the observed line width is likely due to Keplerian motion in the inner disk (condition 4). Hence, the radius derived via the Keplerian relation, $(1.9^{+1.9}_{-1.5})\times10^5\sin^2 i$ km, should be equivalent to the disk truncation radius $R_{\rm M}$. In this case, the accretion system must be nearly face-on to the observer, with $i=4^{\circ}\--13^\circ$ to match $R_{\rm M}=2000$ km (or $R_{\rm co}$). Otherwise, we must consider a scenario with a pair of unknown reflecting structures located at the outer regions of the disk, irrelevant to the inner disk wall, or one in which X-2 is not in the spin equilibrium, in which case the true $R_{\rm M}$ value is indeed significantly larger than $R_{\rm co}$. The former may conflict with the size constraint derived by condition 4, and we do not have a good physical mechanism for generating such localized structures in the disk. In the latter case, X-2 is expected to be in the propeller regime and X-ray dim, which is inconsistent with the observation. Therefore, we conclude that the scenario with a small inclination angle fits naturally with the observational constraints.

The binary parameters in Table \ref{tab:psearch} provide the binary mass function $f=4\pi^2 (a\sin i)^3/GP_{\rm orb}^2 \sim 1.8 M_{\rm \odot}$. By assuming a reasonable range for the neutron star mass ($1M_{\rm \odot}\--2 M_{\rm \odot}$) and a high-mass ($10M_{\rm \odot}\--20 M_{\rm \odot}$) companion star, the function provides a restriction to the orbital inclination angle as $25^{\circ}\--40^{\circ}$. Thus, if we assume that the line width reflects the inner disk Keplerian velocity, the result requires the orbital motion plane to be misaligned from the accretion flow by $20^{\circ}\--30^{\circ}$. Such disagreements between the disk inclination angle and the binary orbit have been reported in several binary systems (e.g., \citealt{connors2019, miller-jones2019}). Therefore, we consider it feasible that X-2 has a similar misorientation.

\begin{figure}
    \centering
    \includegraphics[width=\columnwidth]{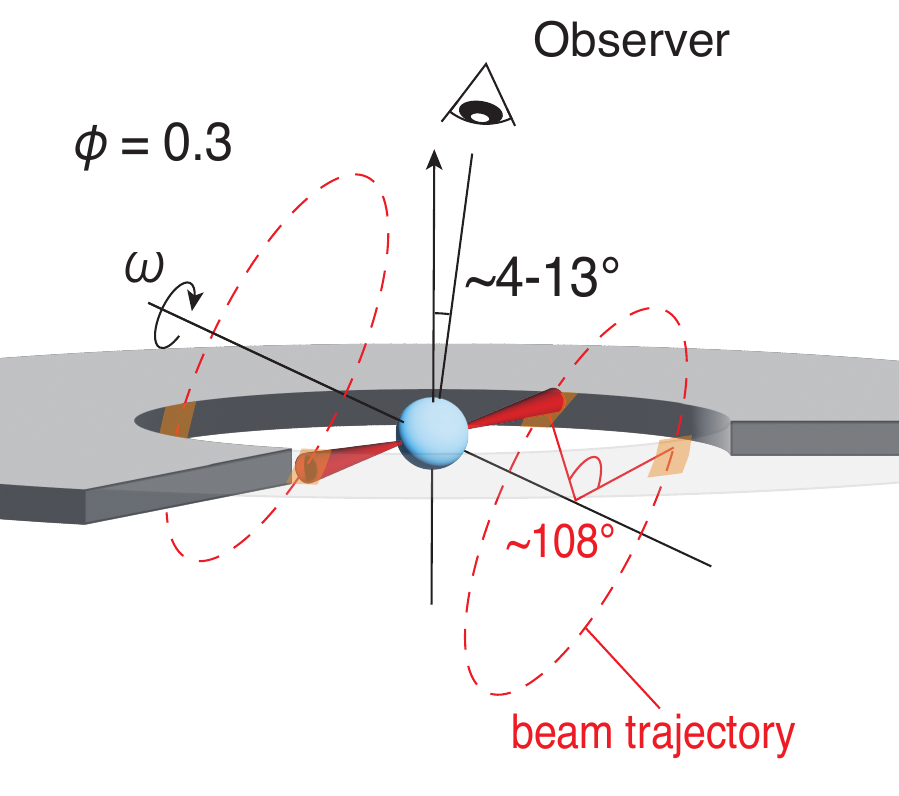}
    \caption{Schematic drawing of a possible accretion geometry of M82 X-2.}
    \label{fig:schem_view}
\end{figure}
Figure \ref{fig:schem_view} is a schematic drawing of what we propose as one of the possible accretion geometries that may fulfill the conditions discussed so far. A geometrically thin accretion disk (gray) is truncated at $R_{\rm M}\sim2000$ km before being inflated by the radiation pressure, and the observer's line of sight is nearly face-on ($i\sim 4^\circ\--13^\circ$) to it. The figure shows a snapshot at $\phi=0.3$ of the rotation phase, namely one of the intensity peaks of the Fe K$\alpha$ line band. Thus, the polar beams, shown as red cones, periodically illuminate the inner disk regions (drawn in orange) as they rotate along the trajectory indicated by the red dashed lines. Since the peak-to-peak distance of the Fe K$\alpha$ band pulse profile is not $180^{\circ}$, it is unlikely that the bimodal shape is due to the illumination by the two opposing sides of the polar beam (as described in Section \ref{sec:lmits_from_pulse_profile}), but each side of the beam is illuminating the inner disk twice with a $108^\circ$ angle interval in a single rotation. This can be realized if the rotation axis of the neutron star is highly inclined from the disk axis, as shown in Figure \ref{fig:schem_view}. In this way, as an addition, each side of the polar beam illuminates the opposite side of the inner disk simultaneously, resulting in a broadening of the line as the present result by observing a summation of the Fe lines from each side, whose central energies are symmetrically blue/red-shifted due to the Kepler motion. Although the physical origin of this high neutron star inclination is unknown, it may be explained by a kick induced at the formation of the neutron star.

We must note that the geometry shown in Figure \ref{fig:schem_view} is one of the possible scenarios that can roughly explain the observational results. For example, we have assumed a neutron star with a dipole magnetic field, which is not necessarily guaranteed, and an alternative explanation may be possible if we consider fields with quadrupole or higher-order components. Furthermore, the physical nature of the asymmetric (and possibly multi-peaked) pulse profile of the continuum band is still unclear. The shape strongly suggests that the accretion column has a rather complex geometry or emission pattern. Accordingly, it may require more sophisticated accretion models and numerical simulations to reproduce the pulsation pattern, and we leave this as future work. 

The observation was conducted under photon-limited conditions due to the configuration with the gate valve closed. Yet, Resolve still provided new insight into the source. Resolve has provided tentative evidence that the neutral Fe K$\alpha$ line may vary in phase with the candidate pulsar rotation period, suggesting that, if confirmed, a fraction of the emission would be attributed to ULXP X-2. The obtained line width, $36^{+60}_{-13}$ eV, is too wide to be explained by turbulent motion on the surface of the companion star, and the variability time scale is too short for the line to originate from the distance or size of the surrounding stellar wind. Therefore, we concluded that the Fe K$\alpha$ line is likely emission from accretion flow in the vicinity of the neutron star, such as the inner region of the accretion disk. Furthermore, Resolve has provided hints of several additional interesting features, such as the possible absorption line feature and emission lines from mildly ionized Fe. Therefore, follow-up observations of M82 X-2 with XRISM are warranted and strongly encouraged.
\begin{acknowledgments}
The authors would like to thank all the members of the  XRISM teams for their devotion to instrumental calibration and spacecraft operation. This research has been supported by JSPS KAKENHI grant numbers 22H00158, 23H04899, 25H00672, and 25K07356. EB acknowledges support by NASA under award number 80GSFC24M0006. ES and SG
acknowledge support by NASA under award number 80NSSC23K0646.
\end{acknowledgments}

%

\vspace{5mm}
\facilities{XRISM}


\software{Adobe illustrator (\url{https://adobe.com/products/illustrator}), ASTROPY \citep{collaboration2018}, HEASOFT (\url{https://heasarc.gsfc.nasa.gov/docs/software/heasoft/}), MATPLOTLIB \citep{hunter2007}, NUMPY \citep{harris2020}, ROOT (\url{https://root.cern/}), and VEUSZ (\url{https://veusz.github.io/})
          }



\appendix

\section{significance contour \label{sec:p_search_significance}}
Figure \ref{fig:search_result} presents the $Z^2_{n}$ heat map around the best estimate parameter shown in table \ref{tab:psearch}. 
\begin{figure}
\includegraphics[width=\columnwidth]
    {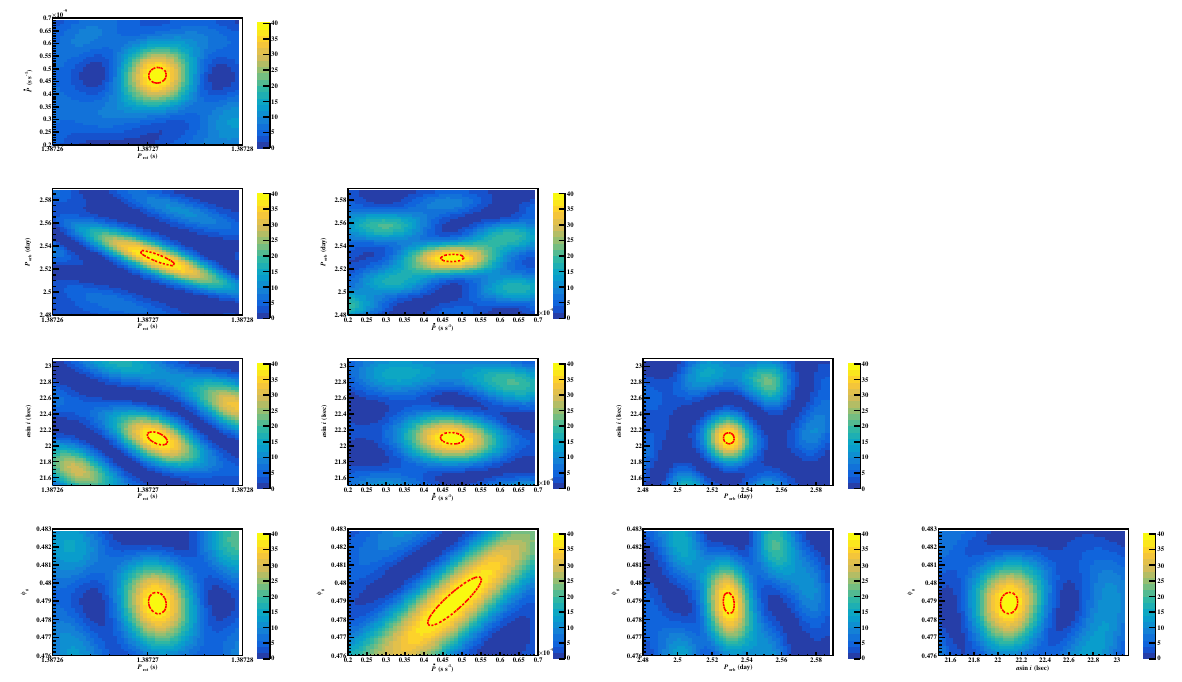}
    \caption{The $Z^2_n$ heat map obtained in analysis done in \ref{sec:pulse_ana}. The color bar represents the $Z^{2}_n$ value. The red dashed lines are significance contours indicating $1\sigma$ confidence level.}
    \label{fig:search_result}
\end{figure}


\bibliography{ULX_related}{}

@article{audard2026,
  title = {A Fast Starburst Wind Consumes Most of the Energy from Supernovae},
  author = {Audard, Marc and Awaki, Hisamitsu and Ballhausen, Ralf and Bamba, Aya and Behar, Ehud and {Boissay-Malaquin}, Rozenn and Brenneman, Laura and Brown, Gregory V. and Corrales, Lia and Costantini, Elisa and Cumbee, Renata and D{\'i}az Trigo, Mar{\'i}a and Done, Chris and Dotani, Tadayasu and Ebisawa, Ken and Eckart, Megan E. and Eckert, Dominique and Eguchi, Satoshi and Enoto, Teruaki and Ezoe, Yuichiro and Foster, Adam and Fujimoto, Ryuichi and Fujita, Yutaka and Fukazawa, Yasushi and Fukushima, Kotaro and Furuzawa, Akihiro and Gallo, Luigi and Garc{\'i}a, Javier A. and Gu, Liyi and Guainazzi, Matteo and Hagino, Kouichi and Hamaguchi, Kenji and Hatsukade, Isamu and Hayashi, Katsuhiro and Hayashi, Takayuki and Hell, Natalie and {Hodges-Kluck}, Edmund and Hornschemeier, Ann and Ichinohe, Yuto and Ishi, Daiki and Ishida, Manabu and Ishikawa, Kumi and Ishisaki, Yoshitaka and Kaastra, Jelle and Kallman, Timothy and Kara, Erin and Katsuda, Satoru and Kanemaru, Yoshiaki and Kelley, Richard and Kilbourne, Caroline and Kitamoto, Shunji and Kobayashi, Shogo and Kohmura, Takayoshi and Kubota, Aya and Leutenegger, Maurice and Loewenstein, Michael and Maeda, Yoshitomo and Markevitch, Maxim and Matsumoto, Hironori and Matsushita, Kyoko and McCammon, Dan and McNamara, Brian and Mernier, Fran{\c c}ois and Miller, Eric D. and Miller, Jon M. and Mitsuishi, Ikuyuki and Mizumoto, Misaki and Mizuno, Tsunefumi and Mori, Koji and Mukai, Koji and Murakami, Hiroshi and Mushotzky, Richard and Nakajima, Hiroshi and Nakazawa, Kazuhiro and Ness, Jan-Uwe and Nobukawa, Kumiko and Nobukawa, Masayoshi and Noda, Hirofumi and Odaka, Hirokazu and Ogawa, Shoji and Ogorzalek, Anna and Okajima, Takashi and Ota, Naomi and Paltani, Stephane and Petre, Robert and Plucinsky, Paul and Porter, Frederick S. and Pottschmidt, Katja and Sato, Kosuke and Sato, Toshiki and Sawada, Makoto and Seta, Hiromi and Shidatsu, Megumi and Simionescu, Aurora and Smith, Randall and Suzuki, Hiromasa and Szymkowiak, Andrew and Takahashi, Hiromitsu and Takeo, Mai and Tamagawa, Toru and Tamura, Keisuke and Tanaka, Takaaki and Tanimoto, Atsushi and Tashiro, Makoto and Terada, Yukikatsu and Terashima, Yuichi and Tsuboi, Yohko and Tsujimoto, Masahiro and Tsunemi, Hiroshi and Tsuru, Takeshi and T{\"u}mer, Ay{\c s}eg{\"u}l and Uchida, Hiroyuki and Uchida, Nagomi and Uchida, Yuusuke and Uchiyama, Hideki and Ueda, Yoshihiro and Uno, Shinichiro and Vink, Jacco and Watanabe, Shin and Williams, Brian J. and Yamada, Satoshi and Yamada, Shinya and Yamaguchi, Hiroya and Yamaoka, Kazutaka and Yamasaki, Noriko and Yamauchi, Makoto and Yamauchi, Shigeo and Yaqoob, Tahir and Yoneyama, Tomokage and Yoshida, Tessei and Yukita, Mihoko and Zhuravleva, Irina and Ampuku, Kazuki and Boettcher, Erin and Grayson, Skylar and Grell, Gabriel and Kosec, Peter and Sasamata, Seiya and Scannapieco, Evan and {XRISM Collaboration}},
  year = 2026,
  month = mar,
  journal = {Nature},
  volume = {651},
  number = {8107},
  pages = {909--913},
  publisher = {Nature Publishing Group},
  issn = {1476-4687},
  doi = {10.1038/s41586-026-10231-1},
  urldate = {2026-03-31},
  copyright = {2026 The Author(s), under exclusive licence to Springer Nature Limited},
  langid = {english}
}

@article{bachetti2014,
  title = {An Ultraluminous {{X-ray}} Source Powered by an Accreting Neutron Star},
  author = {Bachetti, M. and Harrison, F. A. and Walton, D. J. and Grefenstette, B. W. and Chakrabarty, D. and F{\"u}rst, F. and Barret, D. and Beloborodov, A. and Boggs, S. E. and Christensen, F. E. and Craig, W. W. and Fabian, A. C. and Hailey, C. J. and Hornschemeier, A. and Kaspi, V. and Kulkarni, S. R. and Maccarone, T. and Miller, J. M. and Rana, V. and Stern, D. and Tendulkar, S. P. and Tomsick, J. and Webb, N. A. and Zhang, W. W.},
  year = 2014,
  month = oct,
  journal = {Nature},
  volume = {514},
  number = {7521},
  pages = {202--204},
  publisher = {Nature Publishing Group},
  issn = {1476-4687},
  doi = {10.1038/nature13791},
  urldate = {2025-03-11},
  copyright = {2014 Springer Nature Limited},
  langid = {english}
}

@article{bachetti2020,
  title = {All at {{Once}}: {{Transient Pulsations}}, {{Spin-down}}, and a {{Glitch}} from the {{Pulsating Ultraluminous X-Ray Source M82 X-2}}},
  shorttitle = {All at {{Once}}},
  author = {Bachetti, Matteo and Maccarone, Thomas J. and Brightman, Murray and Brumback, McKinley C. and F{\"u}rst, Felix and Harrison, Fiona A. and Heida, Marianne and Israel, Gian Luca and Middleton, Matthew J. and Tomsick, John A. and Webb, Natalie A. and Walton, Dominic J.},
  year = 2020,
  month = mar,
  journal = {The Astrophysical Journal},
  volume = {891},
  number = {1},
  pages = {44},
  publisher = {The American Astronomical Society},
  issn = {0004-637X},
  doi = {10.3847/1538-4357/ab6d00},
  urldate = {2025-03-11},
  langid = {english}
}

@article{bachetti2021,
  title = {Extending the {{Z2n}} and {{H Statistics}} to {{Generic Pulsed Profiles}}},
  author = {Bachetti, Matteo and Pilia, Maura and Huppenkothen, Daniela and Ransom, Scott M. and Curatti, Stefano and Ridolfi, Alessandro},
  year = 2021,
  month = mar,
  journal = {The Astrophysical Journal},
  volume = {909},
  number = {1},
  pages = {33},
  publisher = {The American Astronomical Society},
  issn = {0004-637X},
  doi = {10.3847/1538-4357/abda4a},
  urldate = {2025-03-11},
  langid = {english}
}

@article{bachetti2022,
  title = {Orbital {{Decay}} in {{M82 X-2}}},
  author = {Bachetti, Matteo and Heida, Marianne and Maccarone, Thomas and Huppenkothen, Daniela and Israel, Gian Luca and Barret, Didier and Brightman, Murray and Brumback, McKinley and Earnshaw, Hannah P. and Forster, Karl and F{\"u}rst, Felix and Grefenstette, Brian W. and Harrison, Fiona A. and Jaodand, Amruta D. and Madsen, Kristin K. and Middleton, Matthew and Pike, Sean N. and Pilia, Maura and Poutanen, Juri and Stern, Daniel and Tomsick, John A. and Walton, Dominic J. and Webb, Natalie and Wilms, J{\"o}rn},
  year = 2022,
  month = oct,
  journal = {The Astrophysical Journal},
  volume = {937},
  number = {2},
  pages = {125},
  publisher = {The American Astronomical Society},
  issn = {0004-637X},
  doi = {10.3847/1538-4357/ac8d67},
  urldate = {2025-03-11},
  langid = {english}
}

@article{basko1975,
  title = {Radiative Transfer in a Strong Magnetic Field and Accreting {{X-ray}} Pulsars.},
  author = {Basko, M. M. and Sunyaev, R. A.},
  year = 1975,
  month = sep,
  journal = {Astronomy and Astrophysics},
  volume = {42},
  pages = {311--321},
  issn = {0004-6361},
  urldate = {2025-07-30}
}

@article{basko1976,
  title = {The {{Limiting Luminosity}} of {{Accreting Neutron Stars With Magnetic Fields}}},
  author = {Basko, M. M. and Sunyaev, R. A.},
  year = 1976,
  month = may,
  journal = {Monthly Notices of the Royal Astronomical Society},
  volume = {175},
  number = {2},
  pages = {395--417},
  issn = {0035-8711},
  doi = {10.1093/mnras/175.2.395},
  urldate = {2025-07-30}
}

@article{brightman2020,
  title = {Spectral {{Evolution}} of the {{Ultraluminous X-Ray Sources M82 X-1}} and {{X-2}}},
  author = {Brightman, Murray and Walton, Dominic J. and Xu, Yanjun and Earnshaw, Hannah P. and Harrison, Fiona A. and Stern, Daniel and Barret, Didier},
  year = 2020,
  month = jan,
  journal = {The Astrophysical Journal},
  volume = {889},
  number = {1},
  pages = {71},
  publisher = {The American Astronomical Society},
  issn = {0004-637X},
  doi = {10.3847/1538-4357/ab629a},
  urldate = {2025-03-11},
  langid = {english}
}

@article{brightman2022,
  title = {An 8.56 {{keV Absorption Line}} in the {{Hyperluminous X-Ray Source}} in {{NGC}} 4045: {{Ultrafast Outflow}} or {{Cyclotron Line}}?},
  shorttitle = {An 8.56 {{keV Absorption Line}} in the {{Hyperluminous X-Ray Source}} in {{NGC}} 4045},
  author = {Brightman, Murray and Kosec, Peter and F{\"u}rst, Felix and Earnshaw, Hannah and Heida, Marianne and Middleton, Matthew J. and Stern, Daniel and Walton, Dominic J.},
  year = 2022,
  month = apr,
  journal = {The Astrophysical Journal},
  volume = {929},
  number = {2},
  pages = {138},
  issn = {0004-637X},
  doi = {10.3847/1538-4357/ac5e37},
  urldate = {2025-08-29},
  langid = {english}
}

@article{buccheri1983,
  title = {Search for Pulsed {$\gamma$}-Ray Emission from Radio Pulsars in the {{COS-B}} Data.},
  author = {Buccheri, R. and Bennett, K. and Bignami, G. F. and Bloemen, J. B. G. M. and Boriakoff, V. and Caraveo, P. A. and Hermsen, W. and Kanbach, G. and Manchester, R. N. and Masnou, J. L. and {Mayer-Hasselwander}, H. A. and {\"O}zel, M. E. and Paul, J. A. and Sacco, B. and Scarsi, L. and Strong, A. W.},
  year = 1983,
  month = dec,
  journal = {Astronomy and Astrophysics},
  volume = {128},
  pages = {245--251},
  issn = {0004-6361},
  urldate = {2025-03-11}
}

@article{bykov2022,
  title = {{{ULX}} Pulsar {{Swift J0243}}.6+6124 Observations with {{{\emph{NuSTAR}}}} : Dominance of Reflected Emission in the Super-{{Eddington}} State},
  shorttitle = {{{ULX}} Pulsar {{Swift J0243}}.6+6124 Observations with {{{\emph{NuSTAR}}}}},
  author = {Bykov, S D and Gilfanov, M R and Tsygankov, S S and Filippova, E V},
  year = 2022,
  month = sep,
  journal = {Monthly Notices of the Royal Astronomical Society},
  volume = {516},
  number = {2},
  pages = {1601--1611},
  issn = {0035-8711, 1365-2966},
  doi = {10.1093/mnras/stac2239},
  urldate = {2025-03-11},
  copyright = {https://academic.oup.com/journals/pages/open\_access/funder\_policies/chorus/standard\_publication\_model},
  langid = {english}
}

@article{carpano2018,
  title = {Discovery of Pulsations from {{NGC}}\,300\,{{ULX1}} and Its Fast Period Evolution},
  author = {Carpano, S and Haberl, F and Maitra, C and Vasilopoulos, G},
  year = 2018,
  month = may,
  journal = {Monthly Notices of the Royal Astronomical Society: Letters},
  volume = {476},
  number = {1},
  pages = {L45-L49},
  issn = {1745-3925},
  doi = {10.1093/mnrasl/sly030},
  urldate = {2025-03-11}
}

@article{collaboration2018,
  title = {The {{Astropy Project}}: {{Building}} an {{Open-science Project}} and {{Status}} of the v2.0 {{Core Package}}*},
  shorttitle = {The {{Astropy Project}}},
  author = {Collaboration, The Astropy and {Price-Whelan}, A. M. and Sip{\H o}cz, B. M. and G{\"u}nther, H. M. and Lim, P. L. and Crawford, S. M. and Conseil, S. and Shupe, D. L. and Craig, M. W. and Dencheva, N. and Ginsburg, A. and VanderPlas, J. T. and Bradley, L. D. and {P{\'e}rez-Su{\'a}rez}, D. and {de Val-Borro}, M. and Contributors), (Primary Paper and Aldcroft, T. L. and Cruz, K. L. and Robitaille, T. P. and Tollerud, E. J. and Committee), (Astropy Coordination and Ardelean, C. and Babej, T. and Bach, Y. P. and Bachetti, M. and Bakanov, A. V. and Bamford, S. P. and Barentsen, G. and Barmby, P. and Baumbach, A. and Berry, K. L. and Biscani, F. and Boquien, M. and Bostroem, K. A. and Bouma, L. G. and Brammer, G. B. and Bray, E. M. and Breytenbach, H. and Buddelmeijer, H. and Burke, D. J. and Calderone, G. and Rodr{\'i}guez, J. L. Cano and Cara, M. and Cardoso, J. V. M. and Cheedella, S. and Copin, Y. and Corrales, L. and Crichton, D. and D'Avella, D. and Deil, C. and Depagne, {\'E}. and Dietrich, J. P. and Donath, A. and Droettboom, M. and Earl, N. and Erben, T. and Fabbro, S. and Ferreira, L. A. and Finethy, T. and Fox, R. T. and Garrison, L. H. and Gibbons, S. L. J. and Goldstein, D. A. and Gommers, R. and Greco, J. P. and Greenfield, P. and Groener, A. M. and Grollier, F. and Hagen, A. and Hirst, P. and Homeier, D. and Horton, A. J. and Hosseinzadeh, G. and Hu, L. and Hunkeler, J. S. and Ivezi{\'c}, {\v Z}. and Jain, A. and Jenness, T. and Kanarek, G. and Kendrew, S. and Kern, N. S. and Kerzendorf, W. E. and Khvalko, A. and King, J. and Kirkby, D. and Kulkarni, A. M. and Kumar, A. and Lee, A. and Lenz, D. and Littlefair, S. P. and Ma, Z. and Macleod, D. M. and Mastropietro, M. and McCully, C. and Montagnac, S. and Morris, B. M. and Mueller, M. and Mumford, S. J. and Muna, D. and Murphy, N. A. and Nelson, S. and Nguyen, G. H. and Ninan, J. P. and N{\"o}the, M. and Ogaz, S. and Oh, S. and Parejko, J. K. and Parley, N. and Pascual, S. and Patil, R. and Patil, A. A. and Plunkett, A. L. and Prochaska, J. X. and Rastogi, T. and Janga, V. Reddy and Sabater, J. and Sakurikar, P. and Seifert, M. and Sherbert, L. E. and {Sherwood-Taylor}, H. and Shih, A. Y. and Sick, J. and Silbiger, M. T. and Singanamalla, S. and Singer, L. P. and Sladen, P. H. and Sooley, K. A. and Sornarajah, S. and Streicher, O. and Teuben, P. and Thomas, S. W. and Tremblay, G. R. and Turner, J. E. H. and Terr{\'o}n, V. and van Kerkwijk, M. H. and {de la Vega}, A. and Watkins, L. L. and Weaver, B. A. and Whitmore, J. B. and Woillez, J. and Zabalza, V. and Contributors), (Astropy},
  year = 2018,
  month = aug,
  journal = {The Astronomical Journal},
  volume = {156},
  number = {3},
  pages = {123},
  publisher = {The American Astronomical Society},
  issn = {1538-3881},
  doi = {10.3847/1538-3881/aabc4f},
  urldate = {2025-08-07},
  langid = {english}
}

@article{connors2019,
  title = {Conflicting {{Disk Inclination Estimates}} for the {{Black Hole X-Ray Binary XTE J1550}}-564},
  author = {Connors, Riley M. T. and Garc{\'i}a, Javier A. and Steiner, James F. and Grinberg, Victoria and Dauser, Thomas and Sridhar, Navin and Gatuzz, Efrain and Tomsick, John and Markoff, Sera B. and Harrison, Fiona},
  year = 2019,
  month = sep,
  journal = {The Astrophysical Journal},
  volume = {882},
  number = {2},
  pages = {179},
  publisher = {The American Astronomical Society},
  issn = {0004-637X},
  doi = {10.3847/1538-4357/ab35df},
  urldate = {2025-08-05},
  langid = {english}
}

@article{dalcanton2009,
  title = {The {{ACS Nearby Galaxy Survey Treasury}}},
  author = {Dalcanton, Julianne J. and Williams, Benjamin F. and Seth, Anil C. and Dolphin, Andrew and Holtzman, Jon and Rosema, Keith and Skillman, Evan D. and Cole, Andrew and Girardi, L{\'e}o and Gogarten, Stephanie M. and Karachentsev, Igor D. and Olsen, Knut and Weisz, Daniel and Christensen, Charlotte and Freeman, Ken and Gilbert, Karoline and Gallart, Carme and Harris, Jason and Hodge, Paul and {de Jong}, Roelof S. and Karachentseva, Valentina and Mateo, Mario and Stetson, Peter B. and Tavarez, Maritza and Zaritsky, Dennis and Governato, Fabio and Quinn, Thomas},
  year = 2009,
  month = jul,
  journal = {The Astrophysical Journal Supplement Series},
  volume = {183},
  pages = {67--108},
  issn = {0067-0049},
  doi = {10.1088/0067-0049/183/1/67},
  urldate = {2025-08-19}
}

@article{debnath2024,
  title = {{{2D}} Unified Atmosphere and Wind Simulations of {{O-type}} Stars},
  author = {Debnath, D. and Sundqvist, J. O. and Moens, N. and Van Der Sijpt, C. and Verhamme, O. and Poniatowski, L. G.},
  year = 2024,
  month = apr,
  journal = {Astronomy \& Astrophysics},
  volume = {684},
  pages = {A177},
  issn = {0004-6361, 1432-0746},
  doi = {10.1051/0004-6361/202348206},
  urldate = {2025-06-13},
  copyright = {https://creativecommons.org/licenses/by/4.0},
  langid = {english}
}

@article{fabbiano1989,
  title = {X {{Rays From Normal Galaxies}}},
  author = {Fabbiano, G.},
  year = 1989,
  month = sep,
  journal = {Annual Review of Astronomy and Astrophysics},
  volume = {27},
  number = {Volume 27, 1989},
  pages = {87--138},
  publisher = {Annual Reviews},
  issn = {0066-4146, 1545-4282},
  doi = {10.1146/annurev.aa.27.090189.000511},
  urldate = {2025-03-11},
  langid = {english}
}

@article{goobar2014,
  title = {{{THE RISE OF SN 2014J IN THE NEARBY GALAXY M82}}},
  author = {Goobar, A. and Johansson, J. and Amanullah, R. and Cao, Y. and Perley, D. A. and Kasliwal, M. M. and Ferretti, R. and Nugent, P. E. and Harris, C. and {Gal-Yam}, A. and Ofek, E. O. and Tendulkar, S. P. and Dennefeld, M. and Valenti, S. and Arcavi, I. and Banerjee, D. P. K. and Venkataraman, V. and Joshi, V. and Ashok, N. M. and Cenko, S. B. and Diaz, R. F. and Fremling, C. and Horesh, A. and Howell, D. A. and Kulkarni, S. R. and Papadogiannakis, S. and Petrushevska, T. and Sand, D. and Sollerman, J. and Stanishev, V. and Bloom, J. S. and Surace, J. and Dupuy, T. J. and Liu, M. C.},
  year = 2014,
  month = mar,
  journal = {The Astrophysical Journal Letters},
  volume = {784},
  number = {1},
  pages = {L12},
  publisher = {The American Astronomical Society},
  issn = {2041-8205},
  doi = {10.1088/2041-8205/784/1/L12},
  urldate = {2025-03-11},
  langid = {english}
}

@article{harris2020,
  title = {Array Programming with {{NumPy}}},
  author = {Harris, Charles R. and Millman, K. Jarrod and {van der Walt}, St{\'e}fan J. and Gommers, Ralf and Virtanen, Pauli and Cournapeau, David and Wieser, Eric and Taylor, Julian and Berg, Sebastian and Smith, Nathaniel J. and Kern, Robert and Picus, Matti and Hoyer, Stephan and {van Kerkwijk}, Marten H. and Brett, Matthew and Haldane, Allan and {del R{\'i}o}, Jaime Fern{\'a}ndez and Wiebe, Mark and Peterson, Pearu and {G{\'e}rard-Marchant}, Pierre and Sheppard, Kevin and Reddy, Tyler and Weckesser, Warren and Abbasi, Hameer and Gohlke, Christoph and Oliphant, Travis E.},
  year = 2020,
  month = sep,
  journal = {Nature},
  volume = {585},
  number = {7825},
  pages = {357--362},
  issn = {1476-4687},
  doi = {10.1038/s41586-020-2649-2},
  urldate = {2025-08-19},
  copyright = {2020 The Author(s)},
  langid = {english}
}

@article{hayashi2024,
  title = {In-Orbit Performance of the {{XMA}} for {{XRISM}}/{{Resolve}}},
  author = {Hayashi, Takayuki and {Boissay-Malaquin}, Rozenn and Tamura, Keisuke and Okajima, Takashi and Eckart, Megan E. and Leutenegger, Maurice A. and Yaqoob, Tahir and Loewenstein, Mike and Kelley, Richard L. and Porter, F. Scott and Kilbourne, Caroline A. and Chiao, Meng P. and Sneiderman, Gary A. and Cumbee, Renata S. and Ishisaki, Yoshitaka and Fujimoto, Ryuichi and Tsujimoto, Masahiro and Sawada, Makoto and Sato, Toshiki and Maeda, Yoshitomo and Ishida, Manabu and Corrales, Lia and Miller, Eric and Tumer, Aysegul and Takahashi, Hiromitsu and Brenneman, Laura and Kanemaru, Yoshiaki and Mizumoto, Misaki and Markevitch, Maxim and Guainazzi, Matteo},
  year = 2024,
  month = aug,
  urldate = {2025-08-19}
}

@article{hunter2007,
  title = {Matplotlib: {{A 2D Graphics Environment}}},
  shorttitle = {Matplotlib},
  author = {Hunter, John D.},
  year = 2007,
  month = may,
  journal = {Computing in Science \& Engineering},
  volume = {9},
  number = {3},
  pages = {90--95},
  issn = {1558-366X},
  doi = {10.1109/MCSE.2007.55},
  urldate = {2025-08-07}
}

@article{ishisaki2025,
  title = {Resolve Instrument Onboard {{XRISM}}: Design, Integration, and Instrument Test Results},
  shorttitle = {Resolve Instrument Onboard {{XRISM}}},
  author = {Ishisaki, Yoshitaka and Kelley, Richard L. and Awaki, Hisamitsu and Balleza, Jesus C. and Barnstable, Kim R. and Bialas, Thomas G. and {Boissay-Malaquin}, Rozenn and Brown, Gregory V. and Canavan, Edgar R. and Cumbee, Renata S. and Carnahan, Timothy M. and Chiao, Meng P. and Comber, Brian J. and Costantini, Elisa and den Herder, Jan-Willem and Dercksen, Johannes and de Vries, Cor P. and DiPirro, Michael J. and Eckart, Megan E. and Ezoe, Yuichiro and Ferrigno, Carlo and Fujimoto, Ryuichi and Gorter, Nathalie and Graham, Steven M. and Grim, Martin and Hartz, Leslie S. and Hayakawa, Ryota and Hayashi, Takayuki and Hell, Natalie and Hoshino, Akio and Ichinohe, Yuto and Ishida, Manabu and Ishikawa, Kumi and James, Bryan L. and Kenyon, Steven J. and Kilbourne, Caroline A. and Kimball, Mark O. and Kitamoto, Shunji and Leutenegger, Maurice A. and Maeda, Yoshitomo and McCammon, Dan and Miko, Joseph J. and Mizumoto, Misaki and Noda, Hirofumi and Okajima, Takashi and Okamoto, Atsushi and Paltani, Stephane and Porter, Frederick S. and Sato, Kosuke and Sato, Toshiki and Sawada, Makoto and Shinozaki, Keisuke and Shipman, Russell and Shirron, Peter J. and Sneiderman, Gary A. and Soong, Yang and Szymkiewicz, Richard and Szymkowiak, Andrew E. and Takei, Yoh and Tamura, Keisuke and Tsujimoto, Masahiro and Uchida, Yuusuke and Wasserzug, Stephen and Witthoeft, Michael C. and Wolfs, Rob and Yamada, Shinya and Yasuda, Susumu},
  year = 2025,
  month = aug,
  journal = {Journal of Astronomical Telescopes, Instruments, and Systems},
  volume = {11},
  number = {4},
  pages = {042023},
  publisher = {SPIE},
  issn = {2329-4124, 2329-4221},
  doi = {10.1117/1.JATIS.11.4.042023},
  urldate = {2025-12-05}
}

@article{iwasawa2023,
  title = {Origin of the Diffuse 4--8 {{keV}} Emission in {{M}} 82},
  author = {Iwasawa, K. and Norman, C. and Gilli, R. and Gandhi, P. and {Per{\'e}z-Torres}, M. A.},
  year = 2023,
  month = jun,
  journal = {Astronomy \& Astrophysics},
  volume = {674},
  pages = {A77},
  publisher = {EDP Sciences},
  issn = {0004-6361, 1432-0746},
  doi = {10.1051/0004-6361/202245548},
  urldate = {2025-03-11},
  copyright = {\copyright{} The Authors 2023},
  langid = {english}
}

@article{kallman2004,
  title = {Photoionization {{Modeling}} and the {{K Lines}} of {{Iron}}},
  author = {Kallman, T. R. and Palmeri, P. and Bautista, M. A. and Mendoza, C. and Krolik, J. H.},
  year = 2004,
  month = dec,
  journal = {The Astrophysical Journal Supplement Series},
  volume = {155},
  number = {2},
  pages = {675},
  publisher = {IOP Publishing},
  issn = {0067-0049},
  doi = {10.1086/424039},
  urldate = {2025-06-12},
  langid = {english}
}

@article{kawashima2012,
  title = {Comptonized {{Photon Spectra}} of {{Supercritical Black Hole Accretion Flows}} with {{Application}} to {{Ultraluminous X-Ray Sources}}},
  author = {Kawashima, T. and Ohsuga, K. and Mineshige, S. and Yoshida, T. and Heinzeller, D. and Matsumoto, R.},
  year = 2012,
  month = jun,
  journal = {The Astrophysical Journal},
  volume = {752},
  pages = {18},
  publisher = {IOP},
  issn = {0004-637X},
  doi = {10.1088/0004-637X/752/1/18},
  urldate = {2025-06-13}
}

@article{kelley2025,
  title = {Resolve Instrument Onboard the {{X-Ray Imaging}} and {{Spectroscopy Mission}}},
  author = {Kelley, Richard L. and Ishisaki, Yoshitaka and Costantini, Elisa and Awaki, Hisamitsu and Balleza, Jesus C. and Barnstable, Kim R. and Bialas, Thomas G. and {Boissay-Malaquin}, Rozenn and Brown, Gregory V. and Canavan, Edgar R. and Timothy M, Carnahan and Chiao, Meng P. and Comber, Brian J. and Cumbee, Renata S. and den Herder, Jan-Willem and Dercksen, Johannes and de Vries, Cor P. and DiPirro, Michael J. and Eckart, Megan E. and Ezoe, Yuichiro and Ferrigno, Carlo and Fujimoto, Ryuichi and Gorter, Nathalie and Graham, Steven M. and Grim, Martin and Hartz, Leslie S. and Hayakawa, Ryota and Hayashi, Takayuki and Hell, Natalie and Ichinohe, Yuto and Ishi, Daiki and Ishida, Manabu and Ishikawa, Kumi and James, Bryan L. and Kanemaru, Yoshiaki and Kenyon, Steven J. and Kilbourne, Caroline A. and Kimball, Mark and Kitamoto, Shunji and Leutenegger, Maurice A. and Maeda, Yoshitomo and McCammon, Dan and McLaughlin, Brian J. and Miko, Joseph J. and van der Meer, Erik and Mizumoto, Misaki and Noda, Hirofumi and Okajima, Takashi and Okamoto, Atsushi and Paltani, Stephane and Porter, Frederick S. and Reichenthal, Lillian S. and Sato, Kosuke and Sato, Toshiki and Sato, Yoichi and Sawada, Makoto and Shinozaki, Keisuke and Shipman, Russell and Shirron, Peter J. and Sneiderman, Gary A. and Soong, Yang and Szymkiewicz, Richard and Szymkowiak, Andrew E. and Takei, Yoh and Takeo, Mai and Tamura, Keisuke and Tsujimoto, Masahiro and Uchida, Yuusuke and Wasserzug, Stephen and Witthoeft, Michael C. and Wolfs, Rob and Yamada, Shinya and Yamasaki, Noriko Y. and Yasuda, Susumu},
  year = 2025,
  month = nov,
  journal = {Journal of Astronomical Telescopes, Instruments, and Systems},
  volume = {11},
  number = {4},
  pages = {042026},
  publisher = {SPIE},
  issn = {2329-4124, 2329-4221},
  doi = {10.1117/1.JATIS.11.4.042026},
  urldate = {2025-12-05}
}

@article{kobayashi2019,
  title = {A New Possible Accretion Scenario for Ultra-Luminous {{X-ray}} Sources},
  author = {Kobayashi, Shogo B and Nakazawa, K and Makishima, K},
  year = 2019,
  month = oct,
  journal = {Monthly Notices of the Royal Astronomical Society},
  volume = {489},
  number = {1},
  pages = {366--384},
  issn = {0035-8711},
  doi = {10.1093/mnras/stz2139},
  urldate = {2025-03-11}
}

@article{kosec2018,
  title = {Evidence for a Variable {{Ultrafast Outflow}} in the Newly Discovered {{Ultraluminous Pulsar NGC}} 300 {{ULX-1}}},
  author = {Kosec, P and Pinto, C and Walton, D J and Fabian, A C and Bachetti, M and Brightman, M and F{\"u}rst, F and Grefenstette, B W},
  year = 2018,
  month = sep,
  journal = {Monthly Notices of the Royal Astronomical Society},
  volume = {479},
  number = {3},
  pages = {3978--3986},
  issn = {0035-8711},
  doi = {10.1093/mnras/sty1626},
  urldate = {2025-05-22}
}

@article{kosec2018a,
  title = {Searching for Outflows in Ultraluminous {{X-ray}} Sources through High-Resolution {{X-ray}} Spectroscopy},
  author = {Kosec, P. and Pinto, C. and Fabian, A. C. and Walton, D. J.},
  year = 2018,
  month = feb,
  journal = {Monthly Notices of the Royal Astronomical Society},
  volume = {473},
  number = {4},
  pages = {5680--5697},
  issn = {0035-8711},
  doi = {10.1093/mnras/stx2695},
  urldate = {2025-08-29}
}

@article{kosec2021,
  title = {Ionized Emission and Absorption in a Large Sample of Ultraluminous {{X-ray}} Sources},
  author = {Kosec, P and Pinto, C and Reynolds, C S and Guainazzi, M and Kara, E and Walton, D J and Fabian, A C and Parker, M L and Valtchanov, I},
  year = 2021,
  month = dec,
  journal = {Monthly Notices of the Royal Astronomical Society},
  volume = {508},
  number = {3},
  pages = {3569--3588},
  issn = {0035-8711},
  doi = {10.1093/mnras/stab2856},
  urldate = {2025-08-29}
}

@article{liu2024,
  title = {The {{Long-term Spin-down Trend}} of {{Ultraluminous X-Ray Pulsar M82 X-2}}},
  author = {Liu, Jiren},
  year = 2024,
  month = jan,
  journal = {The Astrophysical Journal},
  volume = {961},
  number = {2},
  pages = {196},
  publisher = {The American Astronomical Society},
  issn = {0004-637X},
  doi = {10.3847/1538-4357/ad17c7},
  urldate = {2025-03-11},
  langid = {english}
}

@article{makishima2000,
  title = {The {{Nature}} of {{Ultraluminous Compact X-Ray Sources}} in {{Nearby Spiral Galaxies}}},
  author = {Makishima, Kazuo and Kubota, Aya and Mizuno, Tsunefumi and Ohnishi, Tomohisa and Tashiro, Makoto and Aruga, Yoichi and Asai, Kazumi and Dotani, Tadayasu and Mitsuda, Kazuhisa and Ueda, Yoshihiro and Uno, Shin'ichiro and Yamaoka, Kazutaka and Ebisawa, Ken and Kohmura, Yoshiki and Okada, Kyoko},
  year = 2000,
  month = jun,
  journal = {The Astrophysical Journal},
  volume = {535},
  number = {2},
  pages = {632},
  publisher = {IOP Publishing},
  issn = {0004-637X},
  doi = {10.1086/308868},
  urldate = {2025-03-11},
  langid = {english}
}

@article{miller-jones2019,
  title = {A Rapidly Changing Jet Orientation in the Stellar-Mass Black-Hole System {{V404 Cygni}}},
  author = {{Miller-Jones}, James C. A. and Tetarenko, Alexandra J. and Sivakoff, Gregory R. and Middleton, Matthew J. and Altamirano, Diego and Anderson, Gemma E. and Belloni, Tomaso M. and Fender, Rob P. and Jonker, Peter G. and K{\"o}rding, Elmar G. and Krimm, Hans A. and Maitra, Dipankar and Markoff, Sera and Migliari, Simone and Mooley, Kunal P. and Rupen, Michael P. and Russell, David M. and Russell, Thomas D. and Sarazin, Craig L. and Soria, Roberto and Tudose, Valeriu},
  year = 2019,
  month = may,
  journal = {Nature},
  volume = {569},
  number = {7756},
  pages = {374--377},
  publisher = {Nature Publishing Group},
  issn = {1476-4687},
  doi = {10.1038/s41586-019-1152-0},
  urldate = {2025-08-05},
  copyright = {2019 The Author(s), under exclusive licence to Springer Nature Limited},
  langid = {english}
}

@article{ohsuga2006,
  title = {Two-Dimensional {{Radiation-Hydrodynamic Model}} for {{Limit-Cycle Oscillations}} of {{Luminous Accretion Disks}}},
  author = {Ohsuga, Ken},
  year = 2006,
  month = apr,
  journal = {The Astrophysical Journal},
  volume = {640},
  number = {2},
  pages = {923},
  publisher = {IOP Publishing},
  issn = {0004-637X},
  doi = {10.1086/500184},
  urldate = {2025-06-13},
  langid = {english}
}

@article{pinto2016,
  title = {Resolved Atomic Lines Reveal Outflows in Two Ultraluminous {{X-ray}} Sources},
  author = {Pinto, Ciro and Middleton, Matthew J. and Fabian, Andrew C.},
  year = 2016,
  month = may,
  journal = {Nature},
  volume = {533},
  number = {7601},
  pages = {64--67},
  publisher = {Nature Publishing Group},
  issn = {1476-4687},
  doi = {10.1038/nature17417},
  urldate = {2025-03-11},
  copyright = {2016 Springer Nature Limited},
  langid = {english}
}

@article{pinto2021,
  title = {{{XMM-Newton}} Campaign on the Ultraluminous {{X-ray}} Source {{NGC}} 247 {{ULX-1}}: Outflows},
  shorttitle = {{{XMM-Newton}} Campaign on the Ultraluminous {{X-ray}} Source {{NGC}} 247 {{ULX-1}}},
  author = {Pinto, C and Soria, R and Walton, D J and D'A{\`i}, A and Pintore, F and Kosec, P and Alston, W N and Fuerst, F and Middleton, M J and Roberts, T P and Del~Santo, M and Barret, D and Ambrosi, E and Robba, A and Earnshaw, H and Fabian, A C},
  year = 2021,
  month = aug,
  journal = {Monthly Notices of the Royal Astronomical Society},
  volume = {505},
  number = {4},
  pages = {5058--5074},
  issn = {0035-8711},
  doi = {10.1093/mnras/stab1648},
  urldate = {2025-03-11}
}

@article{porter2024,
  title = {In-Flight Performance of the {{XRISM}}/{{Resolve}} Detector System},
  author = {Porter, Frederick S. and Kilbourne, Caroline A. and Chiao, Meng and Cumbee, Renata and Eckart, Megan E. and Fujimoto, Ryuichi and Ishisaki, Yoshitaka and Kanemaru, Yoshiaki and Kelley, Richard L. and Leutenegger, Maurice and Maeda, Yoshitomo and Mizumoto, Misaki and Sato, Kosuke and Sawada, Makoto and Sneiderman, Gary and Takei, Yoh and Tsujimoto, Masahiro and Uchida, Yuusuke and Watanabe, Tomomi and Yamada, Shinya},
  year = 2024,
  month = aug,
  journal = {Space Telescopes and Instrumentation 2024: Ultraviolet to Gamma Ray},
  volume = {13093},
  pages = {450--466},
  doi = {10.1117/12.3018882},
  urldate = {2025-08-19}
}

@article{poutanen2007,
  title = {Supercritically Accreting Stellar Mass Black Holes as Ultraluminous {{X-ray}} Sources},
  author = {Poutanen, Juri and Lipunova, Galina and Fabrika, Sergei and Butkevich, Alexey G. and Abolmasov, Pavel},
  year = 2007,
  month = may,
  journal = {Monthly Notices of the Royal Astronomical Society},
  volume = {377},
  number = {3},
  pages = {1187--1194},
  issn = {0035-8711},
  doi = {10.1111/j.1365-2966.2007.11668.x},
  urldate = {2025-06-13}
}

@article{poutanen2013,
  title = {A {{REFLECTION MODEL FOR THE CYCLOTRON LINES IN THE SPECTRA OF X-RAY PULSARS}}},
  author = {Poutanen, Juri and Mushtukov, Alexander A. and Suleimanov, Valery F. and Tsygankov, Sergey S. and Nagirner, Dmitrij I. and Doroshenko, Victor and Lutovinov, Alexander A.},
  year = 2013,
  month = oct,
  journal = {The Astrophysical Journal},
  volume = {777},
  number = {2},
  pages = {115},
  publisher = {The American Astronomical Society},
  issn = {0004-637X},
  doi = {10.1088/0004-637X/777/2/115},
  urldate = {2025-07-30},
  langid = {english}
}

@article{prinja1990,
  title = {Terminal {{Velocities}} for a {{Large Sample}} of {{O Stars}}, {{B Supergiants}}, and {{Wolf-Rayet Stars}}},
  author = {Prinja, Raman K. and Barlow, M. J. and Howarth, Ian D.},
  year = 1990,
  month = oct,
  journal = {The Astrophysical Journal},
  volume = {361},
  pages = {607},
  publisher = {IOP},
  issn = {0004-637X},
  doi = {10.1086/169224},
  urldate = {2025-12-09}
}

@article{runacres2002,
  title = {The Outer Evolution of Instability-Generated Structure in Radiatively Driven Stellar Winds},
  author = {Runacres, M. C. and Owocki, S. P.},
  year = 2002,
  month = jan,
  journal = {Astronomy and Astrophysics},
  volume = {381},
  pages = {1015--1025},
  publisher = {EDP},
  issn = {0004-6361},
  doi = {10.1051/0004-6361:20011526},
  urldate = {2025-12-09}
}

@article{shakura1973,
  title = {Black Holes in Binary Systems. {{Observational}} Appearance.},
  author = {Shakura, N. I. and Sunyaev, R. A.},
  year = 1973,
  month = jan,
  journal = {Astronomy and Astrophysics},
  volume = {24},
  pages = {337--355},
  issn = {0004-6361},
  urldate = {2025-06-13}
}

@article{standish1990,
  title = {The Observational Basis for {{JPL}}'s {{DE}} 200, the Planetary Ephemerides of the {{Astronomical Almanac}}},
  author = {Standish, Jr., E. M.},
  year = 1990,
  month = jul,
  journal = {Astronomy and Astrophysics},
  volume = {233},
  pages = {252--271},
  issn = {0004-6361},
  urldate = {2025-03-11}
}

@article{tarter1969,
  title = {The {{Interaction}} of {{X-Ray Sources}} with {{Optically Thin Environments}}},
  author = {Tarter, C. Bruce and Tucker, Wallace H. and Salpeter, Edwin E.},
  year = 1969,
  month = jun,
  journal = {The Astrophysical Journal},
  volume = {156},
  pages = {943},
  publisher = {IOP},
  issn = {0004-637X},
  doi = {10.1086/150026},
  urldate = {2025-07-14}
}

@article{tashiro2025,
  title = {X-{{Ray Imaging}} and {{Spectroscopy Mission}}},
  author = {Tashiro, Makoto and Kelley, Richard and Watanabe, Shin and Maejima, Hironori and Reichenthal, Lillian and Toda, Kenichi and Hartz, Leslie and Santovincenzo, Andrea and Matsushita, Kyoko and Yamaguchi, Hiroya and Petre, Robert and Williams, Brian and Guainazzi, Matteo and Costantini, Elisa and Takei, Yoh and Ishisaki, Yoshitaka and Fujimoto, Ryuichi and {Henegar-Leon}, Joy and Sneiderman, Gary and Tomida, Hiroshi and Mori, Koji and Nakajima, Hiroshi and Terada, Yukikatsu and Holland, Matthew and Loewenstein, Michael and Miller, Eric and Sawada, Makoto and Kallman, Timothy and Kaastra, Jelle and Done, Chris and Enoto, Teruaki and Bamba, Aya and Corrales, Lia and Ueda, Yoshihiro and Kara, Erin and Zhuravleva, Irina and Fujita, Yutaka and Arai, Yoshitaka and Audard, Marc and Awaki, Hisamitsu and Ballhausen, Ralf and Baluta, Chris and Bando, Nobutaka and Behar, Ehud and Bialas, Thomas and {Boissay-Malaquin}, Rozenn and Brenneman, Laura and Brown, Gregory V and Chiao, Meng and Cumbee, Renata and {de~Vries}, Cor and {den~Herder}, Jan-Willem and D{\'i}az~Trigo, Mar{\'i}a and DiPirro, Michael and Dotani, Tadayasu and Carrero, Jacobo Ebrero and Ebisawa, Ken and Eckart, Megan and Eckert, Dominique and Eguchi, Satoshi and Ezoe, Yuichiro and Ferrigno, Carlo and Foster, Adam and Fukazawa, Yasushi and Fukushima, Kotaro and Furuzawa, Akihiro and Gallo, Luigi and Garcia~Martinez, Javier and Gorter, Nathalie and Grim, Martin and Gu, Liyi and Hagino, Kouichi and Hamaguchi, Kenji and Hatsukade, Isamu and Hayashi, Katsuhiro and Hayashi, Takayuki and Hell, Natalie and {Hodges-Kluck}, Edmund and Horiuchi, Takafumi and Hornschemeier, Ann and Hoshino, Akio and Ichinohe, Yuto and Ikuta, Chisato and Iizuka, Ryo and Ishi, Daiki and Ishida, Manabu and Ishihama, Naoki and Ishikawa, Kumi and Ishimura, Kosei and Jaffe, Tess and Katsuda, Satoru and Kanemaru, Yoshiaki and Kenyon, Steven and Kilbourne, Caroline and Kimball, Mark and Kitamoto, Shunji and Kobayashi, Shogo and Kohmura, Takayoshi and Kubota, Aya and Leutenegger, Maurice and Maeda, Yoshitomo and Markevitch, Maxim and Matsumoto, Hironori and Matsuzaki, Keiichi and McCammon, Dan and McLaughlin, Brian and McNamara, Brian and Mernier, Francois and Miko, Joseph and Miller, Jon and Minesugi, Kenji and Mitani, Shinji and Mitsuishi, Ikuyuki and Mizumoto, Misaki and Mizuno, Tsunefumi and Mukai, Koji and Murakami, Hiroshi and Mushotzky, Richard and Nakazawa, Kazuhiro and Natsukari, Chikara and Ness, Jan-Uwe and Nigo, Kenichiro and Nishiyama, Mari and Nobukawa, Kumiko and Nobukawa, Masayoshi and Noda, Hirofumi and Odaka, Hirokazu and Ogawa, Mina and Ogawa, Shoji and Ogorzalek, Anna and Okajima, Takashi and Okamoto, Atsushi and Ota, Naomi and Ozaki, Masanobu and Paltani, Stephane and Plucinsky, Paul and Porter, F Scott and Pottschmidt, Katja and Quero, Jose Antonio and Sasaki, Takahiro and Sato, Kosuke and Sato, Rie and Sato, Toshiki and Sato, Yoichi and Seta, Hiromi and Shida, Maki and Shidatsu, Megumi and Shigeto, Shuhei and Shipman, Russel and Shinozaki, Keisuke and Shirron, Peter and Simionescu, Aurora and Smith, Randall and Soong, Yang and Suzuki, Hiromasa and Szymkowiak, Andrew and Takahashi, Hiromitsu and Takeo, Mai and Tamagawa, Toru and Tamura, Keisuke and Tanaka, Takaaki and Tanimoto, Atsushi and Terashima, Yuichi and Tsuboi, Yohko and Tsujimoto, Masahiro and Tsunemi, Hiroshi and Tsuru, Takeshi and Uchida, Hiroyuki and Uchida, Nagomi and Uchida, Yuusuke and Uchiyama, Hideki and Uno, Shinichiro and Vink, Jacco and Witthoeft, Michael and Wolfs, Rob and Yamada, Satoshi and Yamada, Shinya and Yamaoka, Kazutaka and Yamasaki, Noriko and Yamauchi, Makoto and Yamauchi, Shigeo and Yanagase, Keiichi and Yaqoob, Tahir and Yasuda, Susumu and Yoneyama, Tomokage and Yoshida, Tessei and Yukita, Miohoko},
  year = 2025,
  month = apr,
  journal = {Publications of the Astronomical Society of Japan},
  pages = {psaf023},
  issn = {2053-051X},
  doi = {10.1093/pasj/psaf023},
  urldate = {2025-04-14}
}

@article{terada2025,
  title = {Development of the Timing System for the {{X-ray}} Imaging and Spectroscopy Mission},
  author = {Terada, Yukikatsu and Shidatsu, Megumi and Sawada, Makoto and Kominato, Takashi and Kato, So and Sato, Ryohei and Sakama, Minami and Shioiri, Takumi and Niida, Yuki and Natsukari, Chikara and Tashiro, Makoto S. and Toda, Kenichi and Maejima, Hironori and Hayashi, Katsuhiro and Yoshida, Tessei and Ogawa, Shoji and Kanemaru, Yoshiaki and Hoshino, Akio and Fukushima, Kotaro and Takahashi, Hiromitsu and Nobukawa, Masayoshi and Mizuno, Tsunefumi and Nakazawa, Kazuhiro and Uno, Shin'ichiro and Ebisawa, Ken and Eguchi, Satoshi and Katsuda, Satoru and Kubota, Aya and Ota, Naomi and Tanimoto, Atsushi and Terashima, Yuichi and Tsuboi, Yohko and Uchida, Yuusuke and Uchiyama, Hideki and Yamauchi, Shigeo and Yoneyama, Tomokage and Yamada, Satoshi and Uchida, Nagomi and Watanabe, Shin and Iizuka, Ryo and Sato, Rie and Baluta, Chris and Holland, Matt and Loewenstein, Michael and Miller, Eric D. and Yaqoob, Tahir and Hill, Robert S. and Waddy, Morgan D. and Mekosh, Mark M. and Fox, Joseph B. and Aldoretta, Emily J. and Brewer, Isabella and Mukai, Koji and Hamaguchi, Kenji and Mernier, Fran{\c c}ois and Ogorzalek, Anna and Pottschmidt, Katja and Yukita, Mihoko and Takagi, Toshihiro and Motogami, Yugo and Enoto, Teruaki and Tanaka, Takaaki and Nakamoto, Taichi and Kang, Chulsoo and Miyazaki, Tsuyoshi},
  year = 2025,
  month = apr,
  journal = {Journal of Astronomical Telescopes, Instruments, and Systems},
  volume = {11},
  number = {4},
  pages = {042007},
  publisher = {SPIE},
  issn = {2329-4124, 2329-4221},
  doi = {10.1117/1.JATIS.11.4.042007},
  urldate = {2025-11-05}
}

@article{trumper2013,
  title = {{{AN ACCRETION MODEL FOR THE ANOMALOUS X-RAY PULSAR 4U}} 0142+61},
  author = {Tr{\"u}mper, J. E. and Dennerl, K. and Kylafis, N. D. and Ertan, {\"U}. and Zezas, A.},
  year = 2013,
  month = jan,
  journal = {The Astrophysical Journal},
  volume = {764},
  number = {1},
  pages = {49},
  publisher = {The American Astronomical Society},
  issn = {0004-637X},
  doi = {10.1088/0004-637X/764/1/49},
  urldate = {2025-07-30},
  langid = {english}
}

@article{walton2013,
  title = {X-{{RAY OUTFLOWS AND SUPER-EDDINGTON ACCRETION IN THE ULTRALUMINOUS X-RAY SOURCE HOLMBERG IX X-1}}},
  author = {Walton, D. J. and Miller, J. M. and Harrison, F. A. and Fabian, A. C. and Roberts, T. P. and Middleton, M. J. and Reis, R. C.},
  year = 2013,
  month = jul,
  journal = {The Astrophysical Journal Letters},
  volume = {773},
  number = {1},
  pages = {L9},
  publisher = {The American Astronomical Society},
  issn = {2041-8205},
  doi = {10.1088/2041-8205/773/1/L9},
  urldate = {2025-03-11},
  langid = {english}
}

@article{walton2016,
  title = {{{AN IRON K COMPONENT TO THE ULTRAFAST OUTFLOW IN NGC}} 1313 {{X-1}}},
  author = {Walton, D. J. and Middleton, M. J. and Pinto, C. and Fabian, A. C. and Bachetti, M. and Barret, D. and Brightman, M. and Fuerst, F. and Harrison, F. A. and Miller, J. M. and Stern, D.},
  year = 2016,
  month = jul,
  journal = {The Astrophysical Journal Letters},
  volume = {826},
  number = {2},
  pages = {L26},
  publisher = {The American Astronomical Society},
  issn = {2041-8205},
  doi = {10.3847/2041-8205/826/2/L26},
  urldate = {2025-03-11},
  langid = {english}
}

@article{xiao2024,
  title = {Pulsed {{Iron Line Emission}} from the {{First Galactic Ultraluminous X-Ray Pulsar Swift J0243}}.6+6124},
  author = {Xiao, Y. X. and Xu, Y. J. and Ge, M. Y. and Lu, F. J. and Zhang, S. N. and Zhang, S. and Tao, L. and Qu, J. L. and Wang, P. J. and Kong, L. D. and Tuo, Y. L. and You, Y. and Zhao, S. J. and Peng, J. Q. and Du, Y. F. and Zhang, Y. H. and Ye, W. T.},
  year = 2024,
  month = apr,
  journal = {The Astrophysical Journal},
  volume = {965},
  number = {1},
  pages = {18},
  publisher = {The American Astronomical Society},
  issn = {0004-637X},
  doi = {10.3847/1538-4357/ad24f8},
  urldate = {2025-06-05},
  langid = {english}
}
\bibliographystyle{aasjournal}



\end{document}